\documentclass[a4paper]{article}

\usepackage[super]{natbib}

\usepackage[english]{babel}
\usepackage[utf8]{inputenc}
\usepackage[T1]{fontenc}

\usepackage{tcolorbox}

\usepackage{caption}
\usepackage{listings}
\usepackage{graphicx}
\usepackage[autostyle]{csquotes}

\usepackage[a4paper,top=3cm,bottom=2cm,left=3cm,right=3cm,marginparwidth=1.75cm]{geometry}

\usepackage{authblk}
\usepackage{rotating}
\usepackage{csvsimple}
\usepackage{float}

\begin{document}


\title{\huge Bugs as Features (Part II): A Perspective on Enriching Microbiome-Gut-Brain Axis Analyses}
\author[1, 2, +]{Thomaz F. S. Bastiaanssen}
\author[3*]{Thomas P. Quinn}
\author[4*]{Amy Loughman}

\affil[1]{\footnotesize APC Microbiome Ireland, University College Cork, Ireland.
}
\affil[2]{\footnotesize Department of Anatomy and Neuroscience, University College Cork, Ireland.
}
\affil[3]{\footnotesize Independent Scientist, Geelong, Australia
}
\affil[4]{\footnotesize IMPACT (the Institute for Mental and Physical Health and Clinical Translation), Food and Mood Centre, Deakin University, Australia.
}
\affil[+]{\footnotesize Corresponding author: thomazbastiaanssen@gmail.com}
\affil[*]{\footnotesize Joint senior authors}

\date{}

\Affilfont{\fontsize{4}{4}}

\maketitle

\begin{abstract}
    
The microbiome-gut-brain-axis field is multidisciplinary, benefiting from the expertise of microbiology, ecology, psychiatry, computational biology, and epidemiology amongst other disciplines. As the field matures and moves beyond a basic demonstration of its relevance, it is critical that study design and analysis are robust and foster reproducibility. 

In this companion piece to Bugs as Features (part I), we present techniques from adjacent and disparate fields to enrich and inform the analysis of microbiome-gut-brain-axis data. Emerging techniques built specifically for the microbiome-gut-brain axis are also demonstrated. All of these methods are contextualised to inform several common challenges: how do we establish causality? How can we integrate data from multiple 'omics techniques? How might we account for the dynamicism of host-microbiome interactions?

This perspective is offered to  experienced and emerging microbiome scientists alike, to assist with these questions and others, at the study conception, design, analysis and interpretation stages of research.
\end{abstract}

\newpage

\section{Introduction}
The microbiome-gut-brain axis is informed by biological and epistemological knowledge from many disciplines, spanning microbiology, ecology, psychiatry and others. Similarly, in its analysis, it is strengthened by methods from across the scientific landscape, as well as some truly interdisciplinary approaches developed specifically for the microbiome-gut-brain axis field (see \textbf{Fig. 1}). 

In Part I, we introduced core concepts and foundations of compositional data analysis of the  microbiome-gut-brain axis \citep{part1}. From study design and pre-registration of analysis, to selecting the most suitable diversity metrics, and the options for functional inference. Here in part II, we provide a perspective on how to leverage techniques from other disciplines and future directions for the microbiome-gut-brain axis field. We hope that this mapping of the broader landscape will provide useful navigation from which the reader may explore original sources as per their needs and interests.

One aim of this piece is to provide context for the methods borrowed, adapted and developed from both adjacent and far flung fields, and to aid the reader in appraising their respective strengths and weaknesses for microbiome analysis.

As a guiding principle, we believe that the microbiome-gut-brain axis field has an imperative to become a more reproducible science and to operate from a place of deeper statistical and biological understanding. The techniques described below have been carefully examined and selected to ensure that they are fit to drive the field towards this goal. 

\begin{figure}[hbtp]
\includegraphics[width=\textwidth]{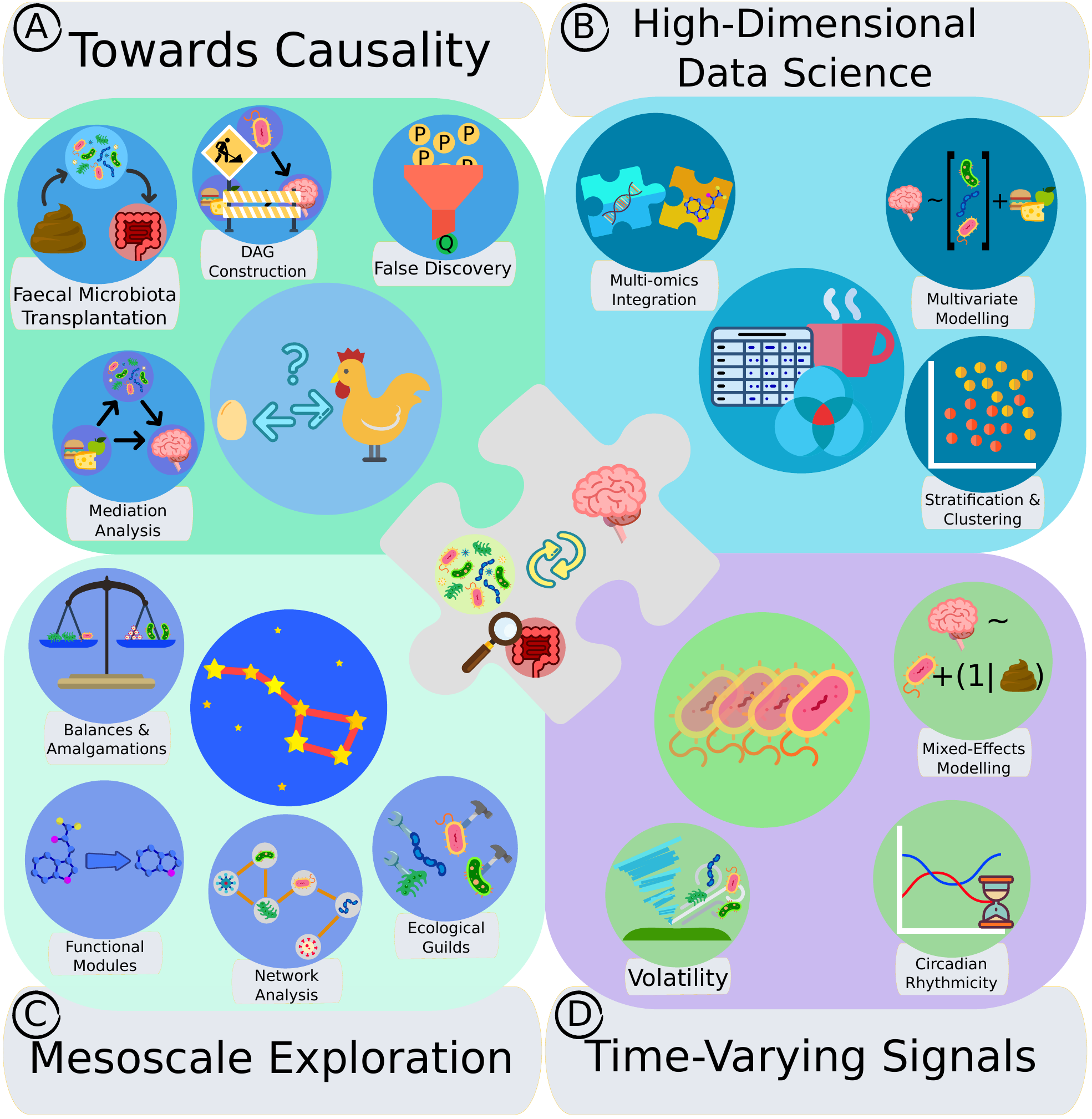}
\centering
\caption{\textbf{Multidisciplinary techniques that enrich the microbiome-gut-brain axis field that are discussed in this perspective.} \textbf{A)} Constructing a directed acyclic graph (DAG) facilitates statistical techniques like causal inference and experimental procedures like Faecal Microbiota Transplantation (FMT) can be used to interrogate causality and directionality. \textbf{B)} Techniques such as multivariate modelling and ordination can be used to analyse and interpret large 'omics data sets. \textbf{C)} 
Higher-order patterns within microbiome data such as interaction networks, functional modules, ecological guilds and amalgamations, called
mesoscale features, are used to ask and answer ecologically relevant questions. \textbf{D)} Microbiome time-series can be analysed using mixed-effect models. Special cases of time-series microbiome data, such as microbial volatility as well as circadian rhythms can be used to ask and answer targeted questions.}
\end{figure}
\newpage

\section{Causality and Uncertainty}
\subsection{Causality and the microbiome}
There has been a growing call for experiments that can establish causality in the microbiome-gut-brain axis field \citep{RN110, bastiaanssen2021microbiota}. Causality is a philosophically and statistically contentious term. Granger causality can be thought of as a pragmatic approach to estimating causality between occurences A and B. In a nutshell, if knowledge of the occurrence A helps predict the occurrence of B, A is said to ``Granger-cause'' B. However, in the case  of complex systems such as the microbiome, where nonlinear dynamics are ubiquitous, Granger causality may not be appropriate \citep{sugihara2012detecting}.  
Historically significant sets of criteria  to establish causality between a microbe and a disease exist, including Koch's postulates \citep{koch1877untersuchungen} and the Bradford Hill criteria \citep{hill1965environment}. Since these are less applicable to the ecosystem approach required for the microbiome-gut-brain-axis, we will not elaborate into these criteria (Though see \textbf{Box 2} on FMT, which leverages experimental design to interrogate causality). Rather, we provide an overview of causality concepts from epidemiology and econometrics that have been applied to the microbiome-gut-brain axis, as well as some important pitfalls.

\subsection{Causal inference analysis}
Causal inference is commonly the underlying motivation for microbiome-gut-brain axis studies, even when it is not being explicitly tested. 
As outlined by \cite{hernan2018c}, being explicit about the causal motivations of (even) an observational analysis ``reduces ambiguity in the scientific question, errors in the data analysis, and excesses in the interpretation of the results''. Rather than avoiding causal language because, as the oft-repeated cautionary tale goes, correlation does not mean causation, Hernán suggests that we instead ask clearer causal questions and improve use of causal inference methods such as adjusting for confounding. There are many occasions where randomised controlled trials that would provide stronger evidence of causality are not feasible, biologically plausible or indeed ethical. The complexity of the gastrointestinal and microbial environments are certainly difficult to replicate completely as interventions in clinical and even pre-clinical trials.

A directed acyclic graph (DAG; \citep{RN56}), or causal diagram, is a useful first step in making explicit the causal hypotheses and underlying assumptions about variables in a study. Creating a DAG serves as a prompt to consider, discuss with colleagues, and design analyses. It is best done at the conception phase of a study, so that it may inform aspects of study design, from the timing of data collection, to the list of potentially confounding variables about which to collect data. We stress here that while DAGs are a helpful tool to ask causal questions, they do not necessarily allow the user to quantify causality from cross-sectional data. Specialized mechanistic follow-up studies remain the gold standard in this regard. In brief, hypothesised relationships between variables are represented by arrows between them, pointing from cause to effect. By convention, causal diagrams point left to right, with exposure variables on the left and outcome variable on the right. Then, add any variables that causally impact the main exposure of interest, or the outcome, using arrows between variables to depict the direction of causality. DAGs must be acyclic, that is variables must not contain feedback loops; relationships between variables must be depicted as unidirectional. This differs from infographics of gut-brain interactions, which are frequently bidirectional as per biological reality. An example DAG was presented in companion piece to this article \citep{part1} and we expand on DAG creation in \textbf{Box 1}. Also see \textbf{Fig 2B}.

%
%

\begin{tcolorbox}
\textbf{BOX 1: Constructing a Directed Acyclic Graph (DAG).}
In an illustrative example of the power of interdisciplinary expertise, 
\cite{RN55} builds on the hybrid field of molecular epidemiology to demonstrate the application of causal inference analysis in 'omics, with the following phases: 
\begin{itemize}
\item[\textbf{A:}] \textbf{Ask specific, detailed research questions.} Build a directed acyclic graph. A DAG may be used to identify the hypothesised relationships between the exposure (microbiome), outcome (e.g., schizophrenia) and potentially confounding variables of interest on the basis of prior knowledge \citep{RN57}. By convention, causal diagrams point left to right, with exposure variables on the left and outcome variable on the right. Then, add any variables that causally impact the main exposure of interest, or the outcome, using arrows between variables to depict the direction of causality. DAGs must be acyclic, that is variables must not contain feedback loops; relationships between variables must be depicted as undirectional. For DAGs with bidirectional relationships, assumptions need to be made as to the dominant direction of action in a given model. 

\item[\textbf{B:}] \textbf{Test the exposure-outcome association.} Run unadjusted association analyses between exposure and outcome. This will require operational definition of each. For example, which microbial diversity metric will be used? How is schizophrenia assessed? Is the exposure-disease association linear or non-linear in form?

\item[\textbf{C:}] \textbf{Consider other variables.} Using the DAG, identify potentially confounding variables. This could include any common causes of exposure and outcome (e.g., cigarette smoking which may affect both the gut microbiome and risk of schizophrenia), and include any proxy measures of unmeasured common causes of both exposure and outcome (e.g., family history of schizophrenia as a proxy for unmeasured genetic factors which could impact both the gut microbiome and risk of schizophrenia) \citep{RN58}. In addition, consider technical or processing variables that might affect measurement precision e.g., microbiome sequencing batch effects \citep{McLarenBias, RN59}. Understanding how each of these causal and non-causal potentially confounding variables associate with the exposure and outcome will provide information to help assess if the putative factor is a mediator, an antecedent, instrumental variable (antecedent of exposure), or a disease consequence. If so, the putative factor is not a confounder and should not be adjusted for. As well as adjusting for confounding, one may include disease determinants that are independent of outcome \citep{RN60}. Too many variables will negatively impact power, so is dimension reduction a possibility? Is there collinearity or redundancy?

\item[\textbf{D:}] \textbf{Build multivariable models.} Consider refining the \textit{a priori} DAG on the basis of the data, adding or removing variables as required, and reporting models that are adjusted on the basis of the original DAG, updated DAG and with any additional processing and precision-enhancing variables that reduce measurement error. Whilst it is desirable to investigate all associations manually, \texttt{DAGitty} software \citep{RN61} does provide identification of ‘minimal adjustment sets’ which can be used to block all non-causal paths, to estimate the total or direct effects between exposure and outcome. Non-causal confounders such as batch effects do not fit strictly within this tool of causal inference, but comprise an unwanted source of variance that should nonetheless feature in adjusted models \citep{RN62}.

\item[\textbf{E:}] \textbf{Evaluate non-causal and causal explanations.} Interpret the findings of both unadjusted and adjusted models. Consider possible biases, such as measurement and selection bias, and other explanations of effects such as reverse causality. 
\end{itemize}

We stress here that DAGs are a tool to formulate causal questions, specialized mechanistic follow-up experiments remain the gold standard to establish causality in the microbiome. For a definitive guide on constructing a DAG, we refer the reader to an excellent free online course \citep{dag_course} informed by the authors of the corresponding text \citep{hernanwhatif}. There is also a useful free online tool \citep{RN61}, though in reality DAGs can be drawn as a proverbial back of napkin sketch almost as effectively.

\end{tcolorbox}
%
%

\subsubsection{Mediation analysis}
Mediation analysis is used to investigate whether a variable transmits its effect on the outcome through another mediator variable \citep{mackinnon2007mediation}. For example, an effect of diet on host behaviour is well documented, as are effects of diet on the microbiome \citep{logan2014nutritional}. Similarly, the gut microbiome is also known to affect host behaviour \citep{RN8}. If we were to test whether diet could affect host behaviour via its effects on the microbiome, that would require a mediation analysis.
Where mediation explains a relationship, there are two main possibilities: 
\begin{itemize}
    \item Partial mediation refers to the scenario where there is both a direct effect and an indirect (i.e., mediation) effect. e.g. if diet were to both directly affect behaviour and also indirectly affect behaviour by modulating the microbiome - which in turn affects behaviour.
    \item Complete mediation refers to the scenario where - using the above example again - diet affects only the microbiome, which in turn affects behaviour, but diet on its own does not directly affect behaviour.  
\end{itemize}
One recent example of how mediation analysis can be used in the microbiome-gut-brain axis field can be found in the context of the autism, diet and the microbiome \citep{RN53}. The authors convincingly show that alterations in the microbiomes of autistic children can be explained by a restricted diet, a common trait in autistic children. They concluded that since diet can explain the altered microbiome, that altered microbiome does not play a causal role in the occurrence of autism.  
In a letter to Yap et al\cite{RN53}., Morton et al\cite{morton2022decoupling}. argue that their model implicitly assumed the absence of a relationship between diet and the microbiome (e.g., independence), which is known to be untrue. Morton et al\cite{morton2022decoupling}. argue that a more appropriate model would be one where diet affects 1) host phenotype directly and 2) the microbiome, which in turn affects phenotype. Essentially, they argue that the microbiome acts as a partial mediator in this autism example. See \textbf{Fig. 2B} for two miniature DAGs illustrating these two scenarios. 

Several excellent tools exist to perform mediation analysis. The \texttt{mediation} package in R takes standard generalised linear model fits as input \citep{tingley2014mediation}. Also see the primer on how to perform a mediation analysis in R in the supplementary files. 
\paragraph{}
We note that mediation is accompanied by inherently longitudinal assumptions. One presumes that due to the occurrence of some exposure at time 1, a mediating variable is affected at time 2, and the outcome shift as a result is observed at some point in future (time 3). The use of mediation analysis in cross-sectional observational data, though common, is not considered best practice \citep{fairchild2017best}. The reason for this is that it presumes that the causal chain being tested is correct and precludes an examination of a potential alternative temporal order of the variables. This is particularly relevant for variables which are dynamic, such as diet, the microbiome and mental states. So mediation analysis in cross-sectional observational studies are correlational and need to be validated in targeted follow-up studies. Some alternative options with fewer data requirements have been trialled through data simulation \citep{cain2018time}, demonstrating that sequential mediation - when data for the exposure, mediator and outcome are collected only once each, but at least longitudinally and in a meaningful temporal sequence - can provide adequate sensitivity to identifying the presence of mediation. The gold-standard is the resource-intensive multilevel longitudinal mediation, where variables that represent enduring exposures (such as diet) are collected repeatedly and path coefficients between variables are allowed to vary across individuals. This may also be ideal for contexts in which a significant degree of inter-individual variability might be expected (such as in host-microbiome studies). 

Notably, estimating an indirect effect through mediation analysis requires substantially more power than estimating direct effects in traditional analysis. For instance, one popular method to estimate the effect size of a mediation analysis is to multiply the two coefficients (exposure to mediator and mediator to outcome), which will always yield a coefficient than its two components \citep{baron1986moderator}. Also see \textbf{Box 3} on power calculations.

\subsection{Mendelian Randomization and the Microbiome}
In contrast to causal inference, Mendelian randomization is a statistical method from the field of epidemiology, often used to estimate the causal effects of genetic factors on a phenotype in large cohorts \citep{davey2003mendelian, MR_plos, MR_nat}. In a nutshell, Mendelian randomization leverages the fact that genotype is fixed at conception and therefore takes place \textit{before} the manifestation of a phenotype. This, along with other assumptions, allows the researcher to assess causality and directionality of the exposure (genotype) on the outcome (phenotype). 

Recently, Mendelian randomization has been applied to microbiome data in the sense that genotype is replaced with microbiome metagenomic content. Particular care is therefore necessary. Unlike genotype, the microbiome is not fixed at conception but remains in constant flux throughout life (See section 5 on Time-varying signals). While it makes sense to assess the causal effect of host genotype on the microbiome, for instance in the case of a host metabolic disorder altering the host gut and hence the microbiome \citep{sanna2019causal}, it seems much less clear whether taking the microbial metagenome as a fixed exposure is appropriate.


\begin{tcolorbox}
\textbf{BOX 2: Faecal Microbiota Transplantation (FMT).} Faecal Microbiota Transplantation involves transferring the microbiome from a donor to a recipient host, often after the recipient microbiome has been washed out using antibiotics or by inducing diarrhea. The idea behind the procedure is to transfer a microbial ecosystem, and potentially the influence said ecosystem has on its host along with it. FMT has shown promise as a therapy for a wide array of disorders \citep{FMT_therapy}. 

FMT is also used as an experimental procedure to investigate causality in a preclinical setting. Often in preclinical studies, murine models are used as the recipient of human microbiomes. FMT can be a useful tool to establish that a phenotype can be transferred by the microbiome, implying that the microbiome is a causal factor in the development of a phenotype. FMT experiments have been used in studies that provide evidence for the microbiome-gut-brain axis, including in depression and aging \citep{RN109, RN108}. 
While FMT can be a powerful experimental tool, experimental designs involving FMT are non-trivial and have been the subject of valid criticism \citep{RN110}. One such criticism is that observational and experimental units are often conflated in FMT studies. Observational units refers to the number of recipients, whereas experimental units refers to the number of donors. Essentially, the criticism is that in a study where a behavioural phenotype is transferred from a patient donor to twenty mouse recipients, the N-number is not twenty but rather one, as there was only one donor microbiome that transferred the phenotype.

\paragraph{}

Pooling donor faecal microbiome samples comes with drawbacks and should not be considered the default option, for several reasons. First, pooling masks the inter-donor variance of the microbiome, which makes it difficult to trace back and investigate what features of the donor microbiome may have caused a phenotype to be transferred. Second, pooling produces microbiome compositions never found in nature. It is well known that numerous taxa display competitive exclusion, i.e., they never stably appear in the same ecosystem. Pooling can therefore create unstable microbial ecosystems, which may end up in distinct compositional equilibria in the recipients. 
The recommendation is therefore to power the study based on the number of donors rather than number of recipients \citep{RN111}. Currently, inter-recipient variance in microbiome composition is hard to estimate due to the differences in methodology between FMT studies and we recommend taking several recipients per donor to estimate the inter-recipient variance in colonization, which may also be dependent on the donor. 

\paragraph{}

In terms of statistics, we suggest generalized linear mixed models to account for inter-donor variance, using donor ID as a random effect. These types of models can for instance be found in the highly cited R library \texttt{lme4} \citep{RN112}.

\paragraph{}

Strain transmission and engraftment analysis is a relatively novel field of study with applicability to FMT studies as well as non-FMT horizontal microbe transmission studies \citep{ferretti2018mother, podlesny2022metagenomic, valles2022variation, valles2023person}. Confirming whether a strain has been transferred from a donor sample -rather than simply being very similar but unrelated- requires shotgun metagenomics-level resolution, as the 16S rRNA gene alone does not allow for this level of precision (Though c.f. recent developments in full-length 16S sequencing analysis \citep{johnson2019evaluation}). The most recent version of \texttt{MetaPhlAn} \citep{blanco2023extending} and \texttt{StrainPhlAn} comes with a specialized script to estimate microbial strain transmission and engraftment alike. 
\end{tcolorbox}


\section{High dimensional data science}

\subsection{Stratifying and clustering samples}
In some cases, it is necessary to stratify data into clusters; distinct sub-groups based on microbiome signature. Stratification is a common method of defining enterotypes, which are large subgroups based on microbial taxonomic composition. The precise number of true enterotypes, as well as the best way to define them, is still up for debate (though 3-4 enterotypes are often cited \citep{RN80, RN81}). Initial efforts involved calculating a Jensen-Shannon dissimilarity matrix and performing cluster analysis (clusters here corresponding to enterotypes) using the partition around medoids (PAM) approach \citep{cluster_entero}. More recently, studies have employed the Dirichlet Multinomial Mixtures (DMM) approach, a promising technique to estimate enterotypes from the Bayesian school \citep{DMMs}. In brief, the method involves estimating a probability vector for each sample and then estimating whether these vectors came from the same source (i.e. metacommunity, enterotype) or from separate enterotypes. Enterotypes appear to be important constructs because they are related to factors like host health, diet and exercise, despite some known limitations \citep{RN82}. Notably, bacterial load is not easy to estimate using metagenomic techniques like 16S and shotgun (although c.f., \cite{RN83}), but can rather be assessed by pan-bacterial qPCR, or most accurately, using flow-cytometry. Bacterial load is associated with enterotype identity and may bias results \citep{RN80, RN82, RN81}.

Stratifying samples based on feature abundance is a defensible approach under some circumstances, for instance when pursuing functional groups of microbes that might exhibit competitive exclusion. However, it is rarely advisable to stratify samples into subgroups while in the middle of an analysis of when working with data sets comprising only 10s-100s of samples because there are too few samples for validation. It is especially important to validate data-driven stratification, either in a new cohort or in a subsection of withheld data that can be used as a validation set. Spurious strata can frequently arise from technical or biological artifacts, leading enthusiastic researchers on long and fruitless tangents. Clustering algorithms, by design, will cluster, and can even find seemingly impressive clusters among random noise!

\subsection{Multi-omics integration}
The microbiome refers to the collection of microbial genes in a sample. While the present work focuses on this type of data, other 'omics also exist \citep{RN87, RN86, RN85, RN84}. Microbial genetic data provides evidence of microbe presence as well as their functional potential \citep{aguiar2016metagenomics}. Besides metagenomics, the three most common types of 'omics data in microbiome-gut-brain axis studies are: 
\begin{itemize}
    \item Metabolomics: the metabolites and small molecules in a sample. Mass-spectrometry or nuclear magnetic resonance (NMR) spectroscopy are the most common techniques to measure the metabolome. Metabolomics can shed light on the functional consequences of a given microbiome.  
    \item Metatranscriptomics: the sequencing of RNA in a sample. In practice, metatranscriptomics can be thought of as RNAseq on a microbial community rather than a single organism. Metatranscriptomcs can tell us about the transcriptional activity of a microbial community. A microbe may be present and have a certain gene, but it may not be transcribing that gene \citep{metatranscriptome}. 
    \item Metaproteomics: the proteins in a sample. Typically, metaproteomics relies on specialized mass spectrometry techniques to identify proteins and derive their sequences. Metaproteomics go further than metatranscriptomics and tell us whether the transcribed genes are translated to proteins. 
\end{itemize}
There are 3 broad approaches to data integration, treating data sets as either univariate or multivariate. The suffixes \textit{-variable} and \textit{-variate} are often used interchangeably, but they refer to subtly but meaningfully distinct concepts \citep{mallick2017experimental}. In short, \textit{-variate} refers to the structural nature of the data, whereas \textit{-variable} refers to the structure and number of variables in the statistical model (also \textbf{Fig. 2A}):
\begin{itemize}
    \item Univariate-univariate: With two separate multivariate data sets, one can  perform an acceptable analysis using ``simple'' univariate methods by correlating each microbiome feature (e.g., taxa or gene) with individual features in the other data set, one at a time. Metrics like Pearson's and Spearman's Rank correlation coefficients are commonly used for this purpose. Both of these metrics can be thought of as special cases of a linear model, which we particularly recommend as it allow for the inclusion of covariates. 
    \item Univariate-multivariate: Treat one feature from one data set as a dependent variable, and use all features from the other data set as the predictors. By repeating this for each feature, all associations between the data sets are described.
    \item Multivariate-multivariate: Multivariate regression, such as a canonical correlation analysis or redundancy analysis \citep{RN88}, to obtain a single model that associates all features from one data set with all features from the other data set. The mixOmics package provides a user-friendly implementation of multivariate methods for microbiome research \citep{RN90, RN89}. Similarly, neural networks or other machine learning can be used \citep{RN92, RN91, RN93}.
\end{itemize}

For multivariate-multivariate analysis, one compelling method, \texttt{DIABLO} \citep{singh2019diablo}, extends this approach by comparing association networks between phenotypes, focusing on the interactions between two 'omics data tests rather than the values within the two individual data sets. This permits the discovery of patterns not necessarily visible in either of the individual data sets. 
Note also that it is possible to extend any of these approaches to incorporate external information about known relationships \textit{within} the individual data set, or \textit{across} the two data sets (e.g., via a gene ontology database). For example, \textit{joint pathway analysis} takes advantage of existing biological knowledge structures by mapping two 'omics datasets to the same metabolic pathways and then assessing the joint coverage as a readout of pathway enrichment \citep{RN77, pang2021metaboanalyst}. Such knowledge structures could also be leveraged to constrain an analysis to include only feature pairs that are canonically able to interact according to the database, thus potentially preserving power by avoiding unnecessary hypothesis testing, as exemplified by the \texttt{anansi} framework \citep{anansi}.
Whatever approach one uses, analysts should take care to normalize or transform their data appropriately, especially since correlations can yield spurious results when measured for compositional data \citep{RN39, RN94}. As with differential abundance analysis, multiple multivariate tests should always be accompanied by an FDR-adjustment. When null hypothesis testing is not straightforward in multivariate methods, permutations or algorithmic validation (e.g., cross-validation) may be used there instead.

\begin{figure}[hbtp]
\includegraphics[width=\textwidth]{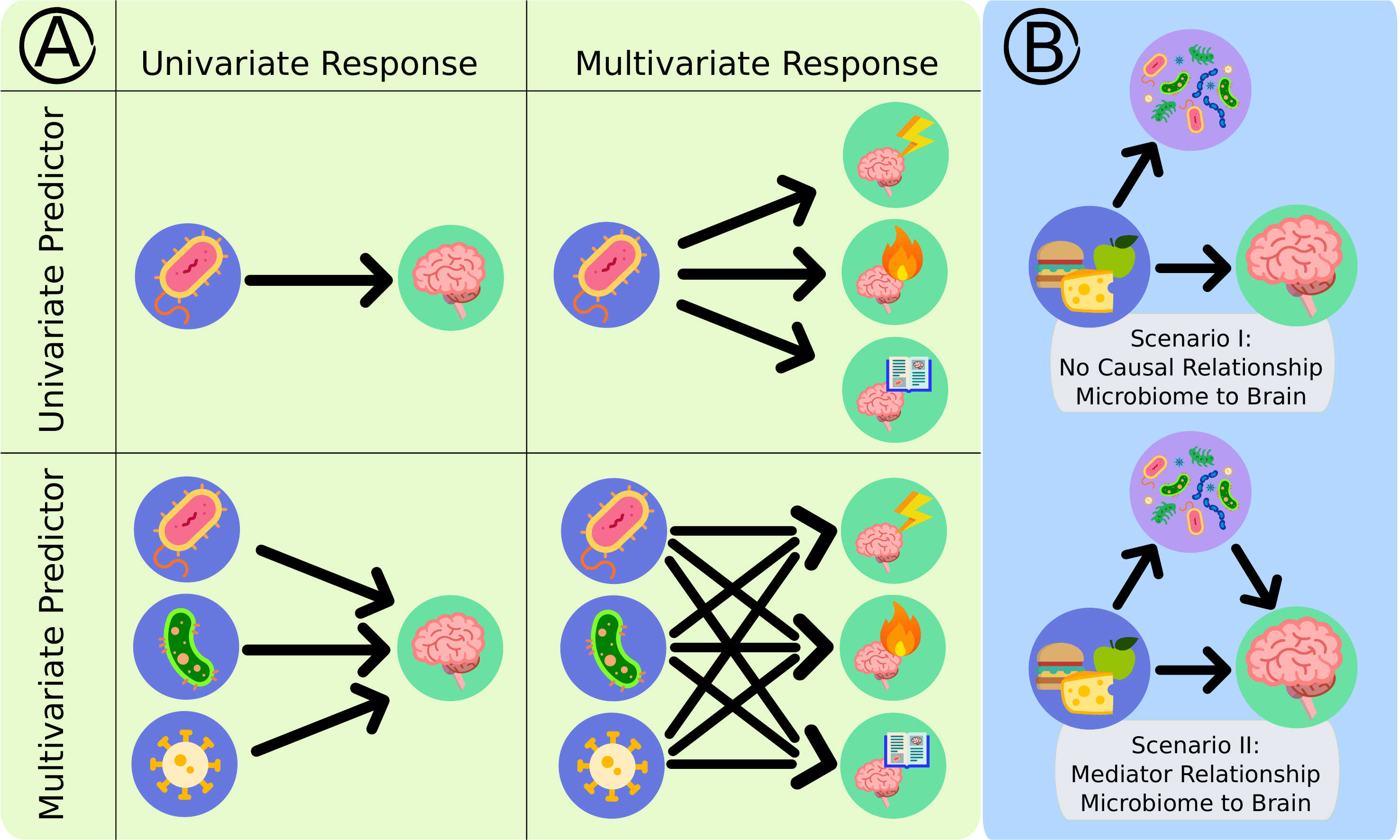}
\centering
\caption{\textbf{Graphical representations of multivariate and mediation analyses using miniature directed acyclic graphs (DAGs).} \textbf{A)} Table illustrating the relationship between models where predictors and responses can be either univariate or multivariate. Predictor variables are on the left of each diagram, whereas response variables are on the right. Univariate predictor variable to multivariate response variables was included for the sake of completion. Practically, a univariate predictor variable to multivariate response variables can approached in the same fashion as multivariate predictor variable to a univariate response variable. \textbf{B)} Two DAGs illustrating the  scenarios from \textbf{2.2.1}. Diet affects the microbiome and diet affects the brain. Mediation analysis helps us ask and answer whether measured associations between the microbiome and the brain are spurious (scenario I) or whether the microbiome affects the brain through diet (scenario II).}
\end{figure}

\newpage
\section{Exploring the mesoscale}
\subsection{Mesoscale features}
Mesoscale features of the microbiome contain information about patterns within parts of a microbiome that can be seen across samples - not necessarily its smallest parts (the microscale), nor about the whole system (the macroscale). Mesoscale analysis focuses on identifying community-level patterns that define the ecosystem(s). This is useful because phenomena in a microbiome may be more readily explained by aggregated patterns in the data rather than by any individual feature. The mesoscale is an important object of study in theoretical ecology \citep{RN95}. These emerging techniques derive microbiome mesoscale features. 
The first three make use of external knowledge. The final two are purely data-driven:

\begin{itemize}
    \item \textbf{Ecological Guilds:} Ecological guilds are taxonomically unrelated but functionally related clusters of microbes that have a shared role in the microbiome (e.g., occupy a common niche). For example, microbial communities across a wide span of environments including soil, the ocean, and the human gut could be assigned to trophic groups based on how they feed on certain substrates and subsequently pass on metabolites to another trophic group \citep{RN96}. While ecological guilds are a promising concept in microbiome science, to our knowledge there are currently no standardized pipelines or databases that can be used to detect and compare ecological guilds across cohorts and experiments (though c.f. \citep{enterosignatures}). Such tools would be welcome additions to the field \citep{RN97, RN98}. 
    
    \item \textbf{Functional Modules:} Functional modules are a list of curated metabolic pathways encoding for processes that are related to a specific aspect of the microbiome. We will consider two classes of functional modules. Gut-Brain Modules cover pathways that are related to gut-brain communication, such as serotonin degradation or histamine production. The complete list of Gut-Brain Modules can be accessed as a table in the supplementary files of \citep{RN99} as well as online (http://raeslab.org/software/gbms.html). Gut-Metabolic Modules, from the same group, encompass metabolic processes in the microbiome. Changes in Gut-Metabolic Modules can indicate a shift in the microbial metabolic environment and thereby in the fitness landscape, thus allowing for microbes with different metabolic features to thrive. The complete list of Gut-Metabolic Modules can be accessed as a table in the supplementary files of the paper that introduced them \citep{RN100} as well as on GitHub (https://github.com/raeslab/GMMs). Functional modules are especially interpretable and help develop hypotheses for future experiments. We note that functional module analysis depends on the availability of a functional abundance table (See the section on functions in the companion piece of this perspective \citep{part1}). 
    
    \item \textbf{Enrichment Analysis:} Differential abundance (DA) analysis is first performed on taxa or genes or other 'omics features, and then a functional database is used to summarize the DA results. In the simplest case, the DA results can be dichotomized into significant or non-significant, and functional status can be dichotomized as present or absent. For each function, one could perform a Fisher exact test (or similar) to test over-enrichment among the significant taxa, genes or features \citep{RN77, RN78}. Gene set enrichment analysis (GSEA) is a popular generalization of this concept, and is commonplace in gene expression analysis \citep{RN78}.

    \item \textbf{Network analysis:} Network analysis is most often applied to study or visualize associations between microbiome features like taxa or genes. This requires some measure of association. The Pearson’s correlation is the most popular, however, correlations have been shown to yield spurious results when applied to compositional data \citep{RN101}. For this reason, several alternatives have been designed specifically for microbiome data \citep{RN40, RN41, RN39}, also see the discussion in the companion piece to this article on compositionality \citep{part1}. These metrics build upon a log-ratio transformation which makes them more robust to the biases introduced by compositionality \citep{RN36}, though they can still be prone to false positives \citep{RN75}. A recent benchmark of 213 single-cell data sets has shown that proportionality has excellent performance for sparse high-dimensional data like those encountered in microbiome research \citep{RN102}.
    
    \item \textbf{Balance selection and summed log-ratios:} Balance selection and data-driven amalgamation are two new approaches to learning mesoscale features directly from the data. In both cases, the motivation is to find mesoscale features that serve as a biomarker to predict another variable-of-interest. These mesoscale features are unique in that they are defined explicitly as a ratio between groups of taxa, similar to the Firmicutes-to-Bacteroidetes ratio \citep{RN103}. By using a ratio of taxa, any normalization factors would cancel, thus making the method normalization-free. When the groups of taxa are summarized by a geometric mean, the resultant mesoscale feature is called a balance. When they are summarized by a sum, the resultant mesoscale feature is called a summed log-ratio (SLR). Software tools like \texttt{selbal} \citep{RN104}, \texttt{balance} \citep{RN105}, \texttt{amalgam} \citep{RN70}, and \texttt{CoDaCoRe} \citep{RN106} enable analysts to learn mesoscale features in a few lines of code. It is customary to validate the reliability of these features by measuring predictive performance in a withheld test set \citep{RN107}.
\end{itemize}

%
%

\begin{tcolorbox}
\textbf{BOX 3: Power calculations and adjustment for multiple testing.} Power calculations can be daunting but are an important part of figuring out how many samples per group are required to register an effect of a given magnitude. Using power calculations can help avoid two undesirable scenarios. First, it saves us from going through the trouble of running an experiment that would not be able to find any effects that may be present (i.e., an underpowered setup). Second, it saves us from collecting more samples than necessary to test a hypothesis (i.e., an overpowered setup). This second scenario is especially important as part of our ethical commitment to avoid excess animal suffering (i.e., “to Reduce” as per the 3Rs of Animal Welfare), as well as participant burden in human studies. Both underpowered and overpowered studies waste precious resources. 

\paragraph{}
In the case of feature-wise analyses such as differential abundance analysis, microbiome power calculations differ from other power calculations for one big reason: microbiome data is highly dimensional. Usually, the number of features is larger than the sample size. This necessitates an adjustment for multiple testing (e.g., by Bonferroni's correction, the Benjamini-Hochberg (BH) procedure or Storey’s q-value \citep{RN73}) which must be accounted for as part of the power calculation. This is because without an FDR correction and instead just using a traditional p < 0.05 threshold, approximately one in twenty tests would give a false positive result, making the p < 0.05 threshold too lenient of an evidence threshold for most applications. Unless there is a compelling reason not to do so, researchers should always correct for multiple testing in microbiome and other high-dimensional datasets. Also see \cite{RN107} for further discussion on differential abundance testing.
For some adjustments, notably the Benjamini-Hochberg procedure and Storey’s q-value, the adjustment depends on the distribution of p-values, which itself depends on the number of true positives. Thus, to account for adjustment, one must estimate how many features will be differentially abundant between treatment groups. This can be difficult to estimate, making it hard to choose the effective FDR-adjusted $\alpha$ threshold for the power calculations.

\paragraph{}

So how to overcome this challenge? At one extreme, we could consider what sample size is needed to detect a single differentially abundant feature if all other features are unaltered. Let’s say only one microbe out of one hundred tested features is actually different. (Incidentally, in the case that only one microbe increases in abundance, the counted proportion of all the others in the sample would decrease. See the discussion in the Part I companion piece to this article on compositionality \citep{part1}). Then, the adjusted p-value needed to reject the null-hypothesis would be one hundred times lower than the initial $\alpha$. Thus, to obtain a conservative sample size estimate, one should perform a power calculation where the significance threshold $\alpha$ is divided by the number of features expected to be tested (not the number anticipated to be differentially abundant).

At the other extreme, we could consider what sample size is needed to detect a single differentially abundant feature if all features differ between groups. In the case that all individual p-values are lower than the alpha threshold, the BH correction will not adjust any of the p-values to the point that they no longer are under $\alpha$. Thus, to obtain a liberal sample size estimate, one should perform a power calculation where the significance threshold $\alpha$ is not adjusted.
From these two extreme scenarios, we can formulate bounds: For $D$ features, the adjusted $\alpha$ used for power calculations should fall between $\frac{\alpha}{D}$ and $\alpha$, depending on the number of features we expect to be differentially abundant. 

\paragraph{}

In the latter case, we recommend a heuristic:
$\alpha_{adjusted} = \alpha \times \frac{M}{D}$
where $D$ is the total number of features tested, and $M$ is the total number of features expected to differ significantly between the groups. This adjusted $alpha$ can then be used for power calculations as done under ordinary circumstances. Once a study is appropriately powered for differential abundance analysis, it seems reasonable to assume that, as a rule-of-thumb, the study is also appropriately powered for testing differences in alpha and beta diversity. Though, to be safe, we recommend that analysts add the total number of planned alpha and beta diversity analyses to $D$ in the formula above.
\end{tcolorbox}

%
%

\section{Time-varying signals}
While 16S and shotgun sequencing only allow for \textit{snapshot} measurements of the microbiome, in reality, microbiomes are dynamic ecosystems in constant flux. In order to account for this, it has become more common for studies to include multiple (repeated) measures of the microbiome. However, time-series analysis necessitates special considerations \citep{kodikara2022statistical,bokulich2018q2}. 

\subsection{Statistical considerations with time-varying data analysis}
Time-series data, where the same microbiomes are sampled repeatedly, intrinsically break the assumption of independence between samples that many statistical tests rely on. Mixed-effects models are well-equipped to handle this type of data, using the resampled microbiome as a random effect. The well-documented and widely used \texttt{lme4} package in R provides an excellent framework for this \citep{RN112}. More specialized microbiome tools such as \texttt{MaAsLin2} \cite{MAASLIN2} are also available. 
 
A recent study on the temporal variation of the microbiome estimates that inter-individual variation is smaller than intra-individual variation \citep{RN72}. Taking several microbiome measurements over time may therefore be necessary to increase power to detect group differences. Another approach to deal with this high intra-individual variation is to include microbial variance in the model \citep{martino2021context}. This allows investigation of whether microbial variability itself is associated to the phenotype of interest. The idea that microbial variance rather than abundance can be informative for a phenotype is core to the idea of volatility.  

\subsection{Volatility}
The microbiome is a dynamic ecosystem that undergoes constant change. The degree of change in the microbiome over time is called volatility, which is inversely related to stability. The term was first coined during the early days of the Human Microbiome Project in the context of instability \citep{weinstock2011volatile} and was soon thereafter used to describe the degree of change in the microbiome between two timepoints \citep{goodrich2014conducting}. It can be helpful to think of volatility as a change in sample diversity (alpha or beta) over time. In a neutral setting, without intervention, a higher volatility is generally considered to be associated with negative health outcomes \citep{RN71}. One way to calculate volatility is to measure the beta diversity between two or more time points corresponding to the same host. When measuring volatility in this fashion, it is especially useful to choose a beta diversity metric that is also a distance (i.e., follows triangle inequality, like PhiLR or Aitchison distance) so that any comparisons are standardized for all time points. Volatility has been recently shown to differ between enterotypes, indicating that microbiome composition at least partially explains microbiome volatility \citep{RN72}. Because sampling depth is known to affect beta diversity indices, it may be worth subsampling before volatility analysis \citep{park2023impact}. 

\subsection{Circadian rhythms}
Circadian rhythms, or 24-hour biological cycles, are key in maintaining physical and mental health \citep{caliyurt2017role}. The microbiome is an example of a biological system that displays such a 24-hour cycle \citep{thaiss2014transkingdom, liang2015rhythmicity}. 
Typically, models that assess rhythmicity will make use of a sinusoidal model rather than a conventional linear model. Circadian rhythms are a special case of time-varying data, as there is an implicit assumption that microbial taxa will oscillate around a set mean (mesor). Due to the 24-hour period of a circadian rhythm, time of sampling becomes an important source of variance and thus a relevant covariate even when the researcher is not interested investigating circadian rhythms \textit{per se}. 
We recently developed the \texttt{kronos} package in R to analyse circadian rhythms in the microbiome \citep{Kronos}. 

\newpage

\section{Consolidating and looking forward}
As the microbiome-gut-brain axis field continues its maturation, we shift our priorities away from a basic demonstration of relevance and towards formulating and addressing more mechanistic questions. In the final section of this perspective, we briefly look forward to efforts to consolidate findings in the field. 
\subsection{Meta-analyses}
In a nutshell, meta-analyses incorporate outcomes from numerous studies on the same subject in order to estimate a `true effect` based on a weighted summary of the component studies. When planning a meta-analysis of microbiome-gut-brain axis studies, it is particularly important to consider which features to analyse. For instance, it may be preferable to investigate the role of microbial functions in a disorder rather than taxonomy-level data. Additionally, due to large inter-study heterogeneity in methodology, it may not be appropriate to compare reported outcomes from studies at all and a `meta-re-analysis` from raw data may be warranted. This again underlines the importance of making microbiome data publicly available. 
We note and applaud the burgeoning development of meta-analysis methods for microbiome studies such as \texttt{MMUPHin} by \cite{ma2022population} which account for the heterogeneity in pre-processing that precludes standard meta-analysis tools and techniques \citep{duvallet2017meta, chong2020meta, Morton_meta_asd}. A groundswell of attempts at reproducing previous findings and quantitative synthesis of the literature to date will improve the robustness of the field, as it has done in others. 

\subsection{Towards enriching microbiome-gut-brain axis research}
In this Part II of our Bugs as Features Perspective, we have taken you on a tour of both adjacent and far flung topics to enrich contemporary microbiome-gut-brain axis research, from Mendelian randomisation and mediation analysis to numerous ways to explore microbiome patterns of the mesoscale. In combination with the concepts and foundations detailed in Part I, and the corresponding supplementary code tutorial, we have described the key considerations for microbiome-gut-brain axis analysis \citep{part1}. In our opinion, establishing causality, integrating multi-omics data and accounting for the dynamic nature of the microbiome are key. We hope that this perspective has assisted with confident navigation of the microbial landscape. We trust that the increased use of biologically and statistically sound methods such as those described here will improve our understanding of the complex phenomenon known as the microbiome-gut-brain axis.

\section{Acknowledgements}
We would like to thank prof. Anne-Louise Ponsonby for her expert comments on DAGs, Dr. Darren L. Dahly for his insights on statistical analysis and prof. John F. Cryan for his excellent advice. We are grateful for their help and support.

\section{Declarations}
The authors declare no competing interests.

APC Microbiome Ireland is a research centre funded by Science Foundation Ireland (SFI), through the Irish Governments’ national development plan (grant no. 12/RC/2273\_P2).


\begin{thebibliography}{1000}

\bibitem{part1}Bastiaanssen, T., Quinn, T. \& Loughman, A. Bugs as Features (Part I): Concepts and Foundations for the Compositional Data Analysis of the Microbiome-Gut-Brain Axis. (arXiv,2022), https://arxiv.org/abs/2207.12475


\bibitem{RN110}
J.~Walter, A.~M. Armet, B.~B. Finlay, and F.~Shanahan, ``Establishing or
  exaggerating causality for the gut microbiome: lessons from human
  microbiota-associated rodents,'' {\em Cell}, vol.~180, no.~2, pp.~221--232,
  2020.

\bibitem{bastiaanssen2021microbiota}
T.~F.~S. Bastiaanssen and J.~F. Cryan, ``The microbiota-gut-brain axis in
  mental health and medication response: parsing directionality and
  causality,'' {\em International Journal of Neuropsychopharmacology}, vol.~24,
  no.~3, pp.~216--220, 2021.

\bibitem{sugihara2012detecting}
G.~Sugihara, R.~May, H.~Ye, C.-h. Hsieh, E.~Deyle, M.~Fogarty, and S.~Munch,
  ``Detecting causality in complex ecosystems,'' {\em science}, vol.~338,
  no.~6106, pp.~496--500, 2012.

\bibitem{koch1877untersuchungen}
R.~Koch, ``Untersuchungen uber bakterien v. die aetiologie der
  milzbrand-krankheit, begrunder auf die entwicklungegeschichte bacillus
  anthracis,'' {\em Beitrage zur biologie der Pflanzen}, vol.~2, no.~2,
  pp.~277--310, 1877.

\bibitem{hill1965environment}
A.~Bradford~Hill, ``The environment and disease: association or causation?,''
  1965.

\bibitem{hernan2018c}
M.~A. Hern{\'a}n, ``The c-word: scientific euphemisms do not improve causal
  inference from observational data,'' {\em American journal of public health},
  vol.~108, no.~5, pp.~616--619, 2018.

\bibitem{RN56}
T.~J. VanderWeele and J.~M. Robins, ``Directed acyclic graphs, sufficient
  causes, and the properties of conditioning on a common effect,'' {\em
  American journal of epidemiology}, vol.~166, no.~9, pp.~1096--1104, 2007.

\bibitem{RN55}
A.-L. Ponsonby, ``Reflection on modern methods: building causal evidence within
  high-dimensional molecular epidemiological studies of moderate size,'' {\em
  International Journal of Epidemiology}, 2021.

\bibitem{RN57}
T.~J. VanderWeele, M.~A. Hernán, and J.~M. Robins, ``Causal directed acyclic
  graphs and the direction of unmeasured confounding bias,'' {\em Epidemiology
  (Cambridge, Mass.)}, vol.~19, no.~5, p.~720, 2008.

\bibitem{RN58}
T.~J. VanderWeele, ``Principles of confounder selection,'' {\em European
  journal of epidemiology}, vol.~34, no.~3, pp.~211--219, 2019.

\bibitem{McLarenBias}
M.~R. McLaren, A.~D. Willis, and B.~J. Callahan, ``Consistent and correctable
  bias in metagenomic sequencing experiments,'' {\em Elife}, vol.~8, p.~e46923,
  2019.

\bibitem{RN59}
Y.~Wang and K.-A. LêCao, ``Managing batch effects in microbiome data,'' {\em
  Briefings in bioinformatics}, vol.~21, no.~6, pp.~1954--1970, 2020.

\bibitem{RN60}
E.~F. Schisterman, S.~R. Cole, and R.~W. Platt, ``Overadjustment bias and
  unnecessary adjustment in epidemiologic studies,'' {\em Epidemiology
  (Cambridge, Mass.)}, vol.~20, no.~4, p.~488, 2009.

\bibitem{RN61}
J.~Textor, B.~van~der Zander, M.~S. Gilthorpe, M.~Liśkiewicz, and G.~T.
  Ellison, ``Robust causal inference using directed acyclic graphs: the r
  package ‘dagitty’,'' {\em International journal of epidemiology},
  vol.~45, no.~6, pp.~1887--1894, 2016.

\bibitem{RN62}
I.~D. Bross, ``Spurious effects from an extraneous variable,'' {\em Journal of
  chronic diseases}, vol.~19, no.~6, pp.~637--647, 1966.

\bibitem{dag_course}
{Hern{\'a}n}, ``Causal diagrams: Draw your assumptions before your
  conclusions,'' 2023.
\newblock [Online; accessed 28-April-2023].

\bibitem{hernanwhatif}
M.~A. Hern{\'a}n and J.~A. Robins, {\em Causal Inference: What If}.
\newblock Boca Raton: Chapman \& Hall/CRC, 2020.

\bibitem{mackinnon2007mediation}
D.~P. MacKinnon, A.~J. Fairchild, and M.~S. Fritz, ``Mediation analysis,'' {\em
  Annu. Rev. Psychol.}, vol.~58, pp.~593--614, 2007.

\bibitem{logan2014nutritional}
A.~C. Logan and F.~N. Jacka, ``Nutritional psychiatry research: an emerging
  discipline and its intersection with global urbanization, environmental
  challenges and the evolutionary mismatch,'' {\em Journal of Physiological
  Anthropology}, vol.~33, pp.~1--16, 2014.

\bibitem{RN8}
J.~F. Cryan, K.~J. O'Riordan, C.~S. Cowan, K.~V. Sandhu, T.~F.~S. Bastiaanssen,
  M.~Boehme, M.~G. Codagnone, S.~Cussotto, C.~Fulling, and A.~V. Golubeva,
  ``The microbiota-gut-brain axis,'' {\em Physiological reviews}, 2019.

\bibitem{RN53}
C.~X. Yap, A.~K. Henders, G.~A. Alvares, D.~L.~A. Wood, L.~Krause, G.~W. Tyson,
  R.~Restuadi, L.~Wallace, T.~McLaren, N.~K. Hansell, D.~Cleary, R.~Grove,
  C.~Hafekost, A.~Harun, H.~Holdsworth, R.~Jellett, F.~Khan, L.~P. Lawson,
  J.~Leslie, M.~L. Frenk, A.~Masi, N.~E. Mathew, M.~Muniandy, M.~Nothard, J.~L.
  Miller, L.~Nunn, G.~Holtmann, L.~T. Strike, G.~I. de~Zubicaray, P.~M.
  Thompson, K.~L. McMahon, M.~J. Wright, P.~M. Visscher, P.~A. Dawson,
  C.~Dissanayake, V.~Eapen, H.~S. Heussler, A.~F. McRae, A.~J.~O. Whitehouse,
  N.~R. Wray, and J.~Gratten, ``Autism-related dietary preferences mediate
  autism-gut microbiome associations,'' {\em Cell}, 2021.

\bibitem{morton2022decoupling}
J.~T. Morton, S.~M. Donovan, and G.~Taroncher-Oldenburg, ``Decoupling diet from
  microbiome dynamics results in model mis-specification that implicitly annuls
  potential associations between the microbiome and disease phenotypes—ruling
  out any role of the microbiome in autism (yap et al. 2021) likely a premature
  conclusion,'' {\em bioRxiv}, pp.~2022--02, 2022.

\bibitem{tingley2014mediation}
D.~Tingley, T.~Yamamoto, K.~Hirose, L.~Keele, and K.~Imai, ``mediation: R
  package for causal mediation analysis,'' {\em Journal of Statistical
  Software}, vol.~59, no.~5, p.~1–38, 2014.

\bibitem{fairchild2017best}
A.~J. Fairchild and H.~L. McDaniel, ``Best (but oft-forgotten) practices:
  mediation analysis,'' {\em The American journal of clinical nutrition},
  vol.~105, no.~6, pp.~1259--1271, 2017.

\bibitem{cain2018time}
M.~K. Cain, Z.~Zhang, and C.~Bergeman, ``Time and other considerations in
  mediation design,'' {\em Educational and psychological measurement}, vol.~78,
  no.~6, pp.~952--972, 2018.

\bibitem{baron1986moderator}
R.~M. Baron and D.~A. Kenny, ``The moderator--mediator variable distinction in
  social psychological research: Conceptual, strategic, and statistical
  considerations.,'' {\em Journal of personality and social psychology},
  vol.~51, no.~6, p.~1173, 1986.

\bibitem{davey2003mendelian}
G.~Davey~Smith and S.~Ebrahim, ``‘mendelian randomization’: can genetic
  epidemiology contribute to understanding environmental determinants of
  disease?,'' {\em International journal of epidemiology}, vol.~32, no.~1,
  pp.~1--22, 2003.

\bibitem{MR_plos}
S.~A. Gagliano~Taliun and D.~M. Evans, ``Ten simple rules for conducting a
  mendelian randomization study,'' 2021.

\bibitem{MR_nat}
E.~Sanderson, M.~M. Glymour, M.~V. Holmes, H.~Kang, J.~Morrison, M.~R.
  Munaf{\`o}, T.~Palmer, C.~M. Schooling, C.~Wallace, Q.~Zhao, {\em et~al.},
  ``Mendelian randomization,'' {\em Nature Reviews Methods Primers}, vol.~2,
  no.~1, p.~6, 2022.

\bibitem{sanna2019causal}
S.~Sanna, N.~R. van Zuydam, A.~Mahajan, A.~Kurilshikov, A.~Vich~Vila,
  U.~V{\~o}sa, Z.~Mujagic, A.~A. Masclee, D.~M. Jonkers, M.~Oosting, {\em
  et~al.}, ``Causal relationships among the gut microbiome, short-chain fatty
  acids and metabolic diseases,'' {\em Nature genetics}, vol.~51, no.~4,
  pp.~600--605, 2019.

\bibitem{FMT_therapy}
K.~Wortelboer, M.~Nieuwdorp, and H.~Herrema, ``Fecal microbiota transplantation
  beyond clostridioides difficile infections,'' {\em EBioMedicine}, vol.~44,
  pp.~716--729, 2019.

\bibitem{RN109}
J.~R. Kelly, Y.~Borre, C.~O’Brien, E.~Patterson, S.~El~Aidy, J.~Deane, P.~J.
  Kennedy, S.~Beers, K.~Scott, G.~Moloney, A.~E. Hoban, L.~Scott,
  P.~Fitzgerald, P.~Ross, C.~Stanton, G.~Clarke, J.~F. Cryan, and T.~G. Dinan,
  ``Transferring the blues: Depression-associated gut microbiota induces
  neurobehavioural changes in the rat,'' {\em Journal of Psychiatric Research},
  vol.~82, pp.~109--118, 2016.

\bibitem{RN108}
M.~Boehme, K.~E. Guzzetta, T.~F.~S. Bastiaanssen, M.~van~de Wouw, G.~M.
  Moloney, A.~Gual-Grau, S.~Spichak, L.~Olavarría-Ramírez, P.~Fitzgerald, and
  E.~Morillas, ``Microbiota from young mice counteracts selective
  age-associated behavioral deficits,'' {\em Nature Aging}, vol.~1, no.~8,
  pp.~666--676, 2021.

\bibitem{RN111}
C.~E. Gheorghe, N.~L. Ritz, J.~A. Martin, H.~R. Wardill, J.~F. Cryan, and
  G.~Clarke, ``Investigating causality with fecal microbiota transplantation in
  rodents: applications, recommendations and pitfalls,'' {\em Gut microbes},
  vol.~13, no.~1, p.~1941711, 2021.

\bibitem{RN112}
D.~Bates, M.~Mächler, B.~Bolker, and S.~Walker, ``Fitting linear mixed-effects
  models using lme4,'' {\em Journal of Statistical Software}, vol.~67, no.~1,
  pp.~1 -- 48, 2015.

\bibitem{ferretti2018mother}
P.~Ferretti, E.~Pasolli, A.~Tett, F.~Asnicar, V.~Gorfer, S.~Fedi, F.~Armanini,
  D.~T. Truong, S.~Manara, M.~Zolfo, {\em et~al.}, ``Mother-to-infant microbial
  transmission from different body sites shapes the developing infant gut
  microbiome,'' {\em Cell host \& microbe}, vol.~24, no.~1, pp.~133--145, 2018.

\bibitem{podlesny2022metagenomic}
D.~Podlesny, C.~Arze, E.~D{\"o}rner, S.~Verma, S.~Dutta, J.~Walter, and W.~F.
  Fricke, ``Metagenomic strain detection with samestr: identification of a
  persisting core gut microbiota transferable by fecal transplantation,'' {\em
  Microbiome}, vol.~10, no.~1, pp.~1--15, 2022.

\bibitem{valles2022variation}
M.~Valles-Colomer, R.~Bacigalupe, S.~Vieira-Silva, S.~Suzuki, Y.~Darzi, R.~Y.
  Tito, T.~Yamada, N.~Segata, J.~Raes, and G.~Falony, ``Variation and
  transmission of the human gut microbiota across multiple familial
  generations,'' {\em Nature Microbiology}, vol.~7, no.~1, pp.~87--96, 2022.

\bibitem{valles2023person}
M.~Valles-Colomer, A.~Blanco-M{\'\i}guez, P.~Manghi, F.~Asnicar, L.~Dubois,
  D.~Golzato, F.~Armanini, F.~Cumbo, K.~D. Huang, S.~Manara, {\em et~al.},
  ``The person-to-person transmission landscape of the gut and oral
  microbiomes,'' {\em Nature}, pp.~1--11, 2023.

\bibitem{johnson2019evaluation}
J.~S. Johnson, D.~J. Spakowicz, B.-Y. Hong, L.~M. Petersen, P.~Demkowicz,
  L.~Chen, S.~R. Leopold, B.~M. Hanson, H.~O. Agresta, M.~Gerstein, {\em
  et~al.}, ``Evaluation of 16s rrna gene sequencing for species and
  strain-level microbiome analysis,'' {\em Nature communications}, vol.~10,
  no.~1, p.~5029, 2019.

\bibitem{blanco2023extending}
A.~Blanco-M{\'\i}guez, F.~Beghini, F.~Cumbo, L.~J. McIver, K.~N. Thompson,
  M.~Zolfo, P.~Manghi, L.~Dubois, K.~D. Huang, A.~M. Thomas, {\em et~al.},
  ``Extending and improving metagenomic taxonomic profiling with
  uncharacterized species using metaphlan 4,'' {\em Nature Biotechnology},
  pp.~1--12, 2023.

\bibitem{RN80}
P.~I. Costea, F.~Hildebrand, A.~Manimozhiyan, F.~Bäckhed, M.~J. Blaser, F.~D.
  Bushman, W.~M. De~Vos, S.~D. Ehrlich, C.~M. Fraser, and M.~Hattori,
  ``Enterotypes in the landscape of gut microbial community composition,'' {\em
  Nature Microbiology}, vol.~3, pp.~8--16, 2018.

\bibitem{RN81}
D.~Vandeputte, G.~Kathagen, K.~D’hoe, S.~Vieira-Silva, M.~Valles-Colomer,
  J.~Sabino, J.~Wang, R.~Y. Tito, L.~De~Commer, and Y.~Darzi, ``Quantitative
  microbiome profiling links gut community variation to microbial load,'' {\em
  Nature}, vol.~551, p.~507, 2017.

\bibitem{cluster_entero}
M.~Arumugam, J.~Raes, E.~Pelletier, D.~Le~Paslier, T.~Yamada, D.~R. Mende,
  G.~R. Fernandes, J.~Tap, T.~Bruls, J.-M. Batto, {\em et~al.}, ``Enterotypes
  of the human gut microbiome,'' {\em nature}, vol.~473, no.~7346,
  pp.~174--180, 2011.

\bibitem{DMMs}
I.~Holmes, K.~Harris, and C.~Quince, ``Dirichlet multinomial mixtures:
  generative models for microbial metagenomics,'' {\em PloS one}, vol.~7,
  no.~2, p.~e30126, 2012.

\bibitem{RN82}
D.~Knights, T.~L. Ward, C.~E. McKinlay, H.~Miller, A.~Gonzalez, D.~McDonald,
  and R.~Knight, ``Rethinking “enterotypes”,'' {\em Cell host \& microbe},
  vol.~16, no.~4, pp.~433--437, 2014.

\bibitem{RN83}
G.~N.~F. Cruz, A.~P. Christoff, and L.~F.~V. de~Oliveira, ``Equivolumetric
  protocol generates library sizes proportional to total microbial load in 16s
  amplicon sequencing,'' {\em Frontiers in microbiology}, vol.~12, p.~425,
  2021.

\bibitem{RN87}
J.~Lloyd-Price, C.~Arze, A.~N. Ananthakrishnan, M.~Schirmer, J.~Avila-Pacheco,
  T.~W. Poon, E.~Andrews, N.~J. Ajami, K.~S. Bonham, C.~J. Brislawn, D.~Casero,
  H.~Courtney, A.~Gonzalez, T.~G. Graeber, A.~B. Hall, K.~Lake, C.~J. Landers,
  H.~Mallick, D.~R. Plichta, M.~Prasad, G.~Rahnavard, J.~Sauk, D.~Shungin,
  Y.~Vázquez-Baeza, R.~A. White, J.~Bishai, K.~Bullock, A.~Deik, C.~Dennis,
  J.~L. Kaplan, H.~Khalili, L.~J. McIver, C.~J. Moran, L.~Nguyen, K.~A. Pierce,
  R.~Schwager, A.~Sirota-Madi, B.~W. Stevens, W.~Tan, J.~J. ten Hoeve,
  G.~Weingart, R.~G. Wilson, V.~Yajnik, J.~Braun, L.~A. Denson, J.~K. Jansson,
  R.~Knight, S.~Kugathasan, D.~P.~B. McGovern, J.~F. Petrosino, T.~S.
  Stappenbeck, H.~S. Winter, C.~B. Clish, E.~A. Franzosa, H.~Vlamakis, R.~J.
  Xavier, C.~Huttenhower, and I.~Investigators, ``Multi-omics of the gut
  microbial ecosystem in inflammatory bowel diseases,'' {\em Nature}, vol.~569,
  no.~7758, pp.~655--662, 2019.

\bibitem{RN86}
A.~Smolinska, D.~I. Tedjo, L.~Blanchet, A.~Bodelier, M.~J. Pierik, A.~A.~M.
  Masclee, J.~Dallinga, P.~H.~M. Savelkoul, D.~M. A.~E. Jonkers, J.~Penders,
  and F.-J. van Schooten, ``Volatile metabolites in breath strongly correlate
  with gut microbiome in cd patients,'' {\em Analytica Chimica Acta},
  vol.~1025, pp.~1--11, 2018.

\bibitem{RN85}
Z.-Z. Tang, G.~Chen, Q.~Hong, S.~Huang, H.~M. Smith, R.~D. Shah, M.~Scholz, and
  J.~F. Ferguson, ``Multi-omic analysis of the microbiome and metabolome in
  healthy subjects reveals microbiome-dependent relationships between diet and
  metabolites,'' {\em Frontiers in Genetics}, vol.~10, 2019.

\bibitem{RN84}
S.~Yachida, S.~Mizutani, H.~Shiroma, S.~Shiba, T.~Nakajima, T.~Sakamoto,
  H.~Watanabe, K.~Masuda, Y.~Nishimoto, M.~Kubo, F.~Hosoda, H.~Rokutan,
  M.~Matsumoto, H.~Takamaru, M.~Yamada, T.~Matsuda, M.~Iwasaki, T.~Yamaji,
  T.~Yachida, T.~Soga, K.~Kurokawa, A.~Toyoda, Y.~Ogura, T.~Hayashi,
  M.~Hatakeyama, H.~Nakagama, Y.~Saito, S.~Fukuda, T.~Shibata, and T.~Yamada,
  ``Metagenomic and metabolomic analyses reveal distinct stage-specific
  phenotypes of the gut microbiota in colorectal cancer,'' {\em Nature
  Medicine}, vol.~25, no.~6, pp.~968--976, 2019.

\bibitem{aguiar2016metagenomics}
V.~Aguiar-Pulido, W.~Huang, V.~Suarez-Ulloa, T.~Cickovski, K.~Mathee, and
  G.~Narasimhan, ``Metagenomics, metatranscriptomics, and metabolomics
  approaches for microbiome analysis: supplementary issue: bioinformatics
  methods and applications for big metagenomics data,'' {\em Evolutionary
  Bioinformatics}, vol.~12, pp.~EBO--S36436, 2016.

\bibitem{metatranscriptome}
G.~S. Abu-Ali, R.~S. Mehta, J.~Lloyd-Price, H.~Mallick, T.~Branck, K.~L. Ivey,
  D.~A. Drew, C.~DuLong, E.~Rimm, J.~Izard, {\em et~al.}, ``Metatranscriptome
  of human faecal microbial communities in a cohort of adult men,'' {\em Nature
  microbiology}, vol.~3, no.~3, pp.~356--366, 2018.

\bibitem{mallick2017experimental}
H.~Mallick, S.~Ma, E.~A. Franzosa, T.~Vatanen, X.~C. Morgan, and
  C.~Huttenhower, ``Experimental design and quantitative analysis of microbial
  community multiomics,'' {\em Genome biology}, vol.~18, no.~1, pp.~1--16,
  2017.

\bibitem{RN88}
C.~Meng, O.~A. Zeleznik, G.~G. Thallinger, B.~Kuster, A.~M. Gholami, and A.~C.
  Culhane, ``Dimension reduction techniques for the integrative analysis of
  multi-omics data,'' {\em Briefings in Bioinformatics}, vol.~17, no.~4,
  pp.~628--641, 2016.

\bibitem{RN90}
K.-A. Lê~Cao, D.~Rossouw, C.~Robert-Granié, and P.~Besse, ``A sparse pls for
  variable selection when integrating omics data,'' {\em Statistical
  applications in genetics and biology, molecular}, vol.~7, no.~1, 2008.

\bibitem{RN89}
F.~Rohart, B.~Gautier, A.~Singh, and K.-A. Lê~Cao, ``mixomics: An r package
  for ‘omics feature selection and multiple data integration,'' {\em PLoS
  computational biology}, vol.~13, no.~11, p.~e1005752, 2017.

\bibitem{RN92}
V.~Le, T.~P. Quinn, T.~Tran, and S.~Venkatesh, ``Deep in the bowel: Highly
  interpretable neural encoder-decoder networks predict gut metabolites from
  gut microbiome,'' {\em BMC Genomics}, vol.~21, no.~4, p.~256, 2020.

\bibitem{RN91}
J.~T. Morton, A.~A. Aksenov, L.~F. Nothias, J.~R. Foulds, R.~A. Quinn, M.~H.
  Badri, T.~L. Swenson, M.~W. Van~Goethem, T.~R. Northen, Y.~Vazquez-Baeza,
  M.~Wang, N.~A. Bokulich, A.~Watters, S.~J. Song, R.~Bonneau, P.~C.
  Dorrestein, and R.~Knight, ``Learning representations of microbe–metabolite
  interactions,'' {\em Nature Methods}, vol.~16, no.~12, pp.~1306--1314, 2019.

\bibitem{RN93}
D.~Reiman, B.~T. Layden, and Y.~Dai, ``Mimenet: Exploring microbiome-metabolome
  relationships using neural networks,'' {\em PLoS Computational Biology},
  vol.~17, no.~5, p.~e1009021, 2021.

\bibitem{singh2019diablo}
A.~Singh, C.~P. Shannon, B.~Gautier, F.~Rohart, M.~Vacher, S.~J. Tebbutt, and
  K.-A. L{\^e}~Cao, ``Diablo: an integrative approach for identifying key
  molecular drivers from multi-omics assays,'' {\em Bioinformatics}, vol.~35,
  no.~17, pp.~3055--3062, 2019.

\bibitem{RN77}
J.~Chong, O.~Soufan, C.~Li, I.~Caraus, S.~Li, G.~Bourque, D.~S. Wishart, and
  J.~Xia, ``Metaboanalyst 4.0: towards more transparent and integrative
  metabolomics analysis,'' {\em Nucleic acids research}, vol.~46, no.~W1,
  pp.~W486--W494, 2018.

\bibitem{pang2021metaboanalyst}
Z.~Pang, J.~Chong, G.~Zhou, D.~A. de~Lima~Morais, L.~Chang, M.~Barrette,
  C.~Gauthier, P.-{\'E}. Jacques, S.~Li, and J.~Xia, ``Metaboanalyst 5.0:
  narrowing the gap between raw spectra and functional insights,'' {\em Nucleic
  acids research}, vol.~49, no.~W1, pp.~W388--W396, 2021.

\bibitem{anansi}
T.~F.~S. Bastiaanssen, T.~P. Quinn, and J.~F. Cryan, ``Knowledge-based
  integration of multi-omic datasets with anansi: Annotation-based analysis of
  specific interactions,'' {\em arXiv}, 2023.

\bibitem{RN39}
T.~P. Quinn, M.~F. Richardson, D.~Lovell, and T.~M. Crowley, ``propr: an
  r-package for identifying proportionally abundant features using
  compositional data analysis,'' {\em Scientific reports}, vol.~7, no.~1,
  pp.~1--9, 2017.

\bibitem{RN94}
T.~P. Quinn and I.~Erb, ``Examining microbe–metabolite correlations by linear
  methods,'' {\em Nature Methods}, vol.~18, no.~1, pp.~37--39, 2021.

\bibitem{RN95}
P.~Hogeweg, {\em Multilevel cellular automata as a tool for studying
  bioinformatic processes}, pp.~19--28.
\newblock Springer, 2010.

\bibitem{RN96}
M.~Gralka, R.~Szabo, R.~Stocker, and O.~X. Cordero, ``Trophic interactions and
  the drivers of microbial community assembly,'' {\em Current Biology},
  vol.~30, no.~19, pp.~R1176--R1188, 2020.

\bibitem{enterosignatures}
C.~Frioux, R.~Ansorge, E.~{\"O}zkurt, C.~G. Nedjad, J.~Fritscher, C.~Quince,
  S.~M. Waszak, and F.~Hildebrand, ``Enterosignatures define common bacterial
  guilds in the human gut microbiome,'' {\em Cell Host \& Microbe}, 2023.

\bibitem{RN97}
Y.~Y. Lam, C.~Zhang, and L.~Zhao, ``Causality in dietary
  interventions—building a case for gut microbiota,'' {\em Genome medicine},
  vol.~10, no.~1, pp.~1--3, 2018.

\bibitem{RN98}
L.~Zhao, F.~Zhang, X.~Ding, G.~Wu, Y.~Y. Lam, X.~Wang, H.~Fu, X.~Xue, C.~Lu,
  and J.~Ma, ``Gut bacteria selectively promoted by dietary fibers alleviate
  type 2 diabetes,'' {\em Science}, vol.~359, no.~6380, pp.~1151--1156, 2018.

\bibitem{RN99}
M.~Valles-Colomer, G.~Falony, Y.~Darzi, E.~F. Tigchelaar, J.~Wang, R.~Y. Tito,
  C.~Schiweck, A.~Kurilshikov, M.~Joossens, C.~Wijmenga, S.~Claes,
  L.~Van~Oudenhove, A.~Zhernakova, S.~Vieira-Silva, and J.~Raes, ``The
  neuroactive potential of the human gut microbiota in quality of life and
  depression,'' {\em Nature Microbiology}, 2019.

\bibitem{RN100}
S.~Vieira-Silva, G.~Falony, Y.~Darzi, G.~Lima-Mendez, R.~G. Yunta, S.~Okuda,
  D.~Vandeputte, M.~Valles-Colomer, F.~Hildebrand, and S.~Chaffron,
  ``Species–function relationships shape ecological properties of the human
  gut microbiome,'' {\em Nature microbiology}, vol.~1, no.~8, pp.~1--8, 2016.

\bibitem{RN78}
R.~A. Irizarry, C.~Wang, Y.~Zhou, and T.~P. Speed, ``Gene set enrichment
  analysis made simple,'' {\em Statistical methods in medical research},
  vol.~18, no.~6, pp.~565--575, 2009.

\bibitem{RN101}
D.~Lovell, V.~Pawlowsky-Glahn, J.~J. Egozcue, S.~Marguerat, and J.~Bähler,
  ``Proportionality: a valid alternative to correlation for relative data,''
  {\em PLoS computational biology}, vol.~11, no.~3, p.~e1004075, 2015.

\bibitem{RN40}
J.~Friedman and E.~J. Alm, ``Inferring correlation networks from genomic survey
  data,'' {\em PLoS Computational Biology}, 2012.

\bibitem{RN41}
Z.~D. Kurtz, C.~L. Müller, E.~R. Miraldi, D.~R. Littman, M.~J. Blaser, and
  R.~A. Bonneau, ``Sparse and compositionally robust inference of microbial
  ecological networks,'' {\em PLoS computational biology}, vol.~11, no.~5,
  p.~e1004226, 2015.

\bibitem{RN36}
T.~P. Quinn, I.~Erb, M.~F. Richardson, and T.~M. Crowley, ``Understanding
  sequencing data as compositions: an outlook and review,'' {\em
  Bioinformatics}, vol.~34, no.~16, pp.~2870--2878, 2018.

\bibitem{RN75}
I.~Erb and C.~Notredame, ``How should we measure proportionality on relative
  gene expression data?,'' {\em Theory in Biosciences}, vol.~135, no.~1,
  pp.~21--36, 2016.

\bibitem{RN102}
M.~A. Skinnider, J.~W. Squair, and L.~J. Foster, ``Evaluating measures of
  association for single-cell transcriptomics,'' {\em Nature Methods}, vol.~16,
  no.~5, pp.~381--386, 2019.

\bibitem{RN103}
D.~Mariat, O.~Firmesse, F.~Levenez, V.~Guimarăes, H.~Sokol, J.~Doré,
  G.~Corthier, and J.~Furet, ``The firmicutes/bacteroidetes ratio of the human
  microbiota changes with age,'' {\em BMC microbiology}, vol.~9, no.~1,
  pp.~1--6, 2009.

\bibitem{RN104}
J.~Rivera-Pinto, J.~J. Egozcue, V.~Pawlowsky-Glahn, R.~Paredes,
  M.~Noguera-Julian, and M.~L. Calle, ``Balances: a new perspective for
  microbiome analysis,'' {\em MSystems}, vol.~3, no.~4, pp.~e00053--18, 2018.

\bibitem{RN105}
T.~P. Quinn, ``Visualizing balances of compositional data: a new alternative to
  balance dendrograms,'' {\em F1000Research}, vol.~7, 2018.

\bibitem{RN70}
T.~P. Quinn and I.~Erb, ``Amalgams: data-driven amalgamation for the
  dimensionality reduction of compositional data,'' {\em NAR Genomics and
  Bioinformatics}, vol.~2, no.~4, p.~lqaa076, 2020.

\bibitem{RN106}
E.~Gordon-Rodriguez, T.~P. Quinn, and J.~P. Cunningham, ``Learning sparse
  log-ratios for high-throughput sequencing data,'' {\em Bioinformatics}, 2021.

\bibitem{RN107}
T.~P. Quinn, E.~Gordon-Rodriguez, and I.~Erb, ``A critique of differential
  abundance analysis, and advocacy for an alternative,'' {\em arXiv preprint},
  2021.

\bibitem{RN73}
J.~D. Storey and R.~Tibshirani, ``Statistical significance for genomewide
  studies,'' {\em Proceedings of the National Academy of Sciences}, vol.~100,
  no.~16, pp.~9440--9445, 2003.

\bibitem{kodikara2022statistical}
S.~Kodikara, S.~Ellul, and K.-A. L{\^e}~Cao, ``Statistical challenges in
  longitudinal microbiome data analysis,'' {\em Briefings in Bioinformatics},
  vol.~23, no.~4, p.~bbac273, 2022.

\bibitem{bokulich2018q2}
N.~A. Bokulich, M.~R. Dillon, Y.~Zhang, J.~R. Rideout, E.~Bolyen, H.~Li, P.~S.
  Albert, and J.~G. Caporaso, ``q2-longitudinal: longitudinal and paired-sample
  analyses of microbiome data,'' {\em MSystems}, vol.~3, no.~6, pp.~e00219--18,
  2018.

\bibitem{MAASLIN2}
H.~Mallick, A.~Rahnavard, L.~J. McIver, S.~Ma, Y.~Zhang, L.~H. Nguyen, T.~L.
  Tickle, G.~Weingart, B.~Ren, E.~H. Schwager, {\em et~al.}, ``Multivariable
  association discovery in population-scale meta-omics studies,'' {\em PLoS
  computational biology}, vol.~17, no.~11, p.~e1009442, 2021.

\bibitem{RN72}
D.~Vandeputte, L.~De~Commer, R.~Y. Tito, G.~Kathagen, J.~Sabino, S.~Vermeire,
  K.~Faust, and J.~Raes, ``Temporal variability in quantitative human gut
  microbiome profiles and implications for clinical research,'' {\em Nature
  Communications}, vol.~12, no.~1, p.~6740, 2021.

\bibitem{martino2021context}
C.~Martino, L.~Shenhav, C.~A. Marotz, G.~Armstrong, D.~McDonald,
  Y.~V{\'a}zquez-Baeza, J.~T. Morton, L.~Jiang, M.~G. Dominguez-Bello, A.~D.
  Swafford, {\em et~al.}, ``Context-aware dimensionality reduction deconvolutes
  gut microbial community dynamics,'' {\em Nature biotechnology}, vol.~39,
  no.~2, pp.~165--168, 2021.

\bibitem{weinstock2011volatile}
G.~M. Weinstock, ``The volatile microbiome,'' {\em Genome biology}, vol.~12,
  no.~5, pp.~1--2, 2011.

\bibitem{goodrich2014conducting}
J.~K. Goodrich, S.~C. Di~Rienzi, A.~C. Poole, O.~Koren, W.~A. Walters, J.~G.
  Caporaso, R.~Knight, and R.~E. Ley, ``Conducting a microbiome study,'' {\em
  Cell}, vol.~158, no.~2, pp.~250--262, 2014.

\bibitem{RN71}
T.~F.~S. Bastiaanssen, A.~Gururajan, M.~van~de Wouw, G.~M. Moloney, N.~L. Ritz,
  C.~M. Long-Smith, N.~C. Wiley, A.~B. Murphy, J.~M. Lyte, and F.~Fouhy,
  ``Volatility as a concept to understand the impact of stress on the
  microbiome,'' {\em Psychoneuroendocrinology}, vol.~124, p.~105047, 2021.

\bibitem{park2023impact}
D.~J. Park and A.~M. Plantinga, ``Impact of data and study characteristics on
  microbiome volatility estimates,'' {\em Genes}, vol.~14, no.~1, p.~218, 2023.

\bibitem{caliyurt2017role}
O.~Caliyurt, ``Role of chronobiology as a transdisciplinary field of research:
  Its applications in treating mood disorders,'' {\em Balkan medical journal},
  vol.~34, no.~6, pp.~514--521, 2017.

\bibitem{thaiss2014transkingdom}
C.~A. Thaiss, D.~Zeevi, M.~Levy, G.~Zilberman-Schapira, J.~Suez, A.~C.
  Tengeler, L.~Abramson, M.~N. Katz, T.~Korem, N.~Zmora, {\em et~al.},
  ``Transkingdom control of microbiota diurnal oscillations promotes metabolic
  homeostasis,'' {\em Cell}, vol.~159, no.~3, pp.~514--529, 2014.

\bibitem{liang2015rhythmicity}
X.~Liang, F.~D. Bushman, and G.~A. FitzGerald, ``Rhythmicity of the intestinal
  microbiota is regulated by gender and the host circadian clock,'' {\em
  Proceedings of the National Academy of Sciences}, vol.~112, no.~33,
  pp.~10479--10484, 2015.

\bibitem{Kronos}
T.~F.~S. Bastiaanssen, S.-J. Leigh, G.~S.~S. Tofani, C.~E. Gheorghe, G.~Clarke,
  and J.~F. Cryan, ``Kronos: A computational tool to facilitate biological
  rhythmicity analysis,'' {\em bioRxiv}, 2023.

\bibitem{ma2022population}
S.~Ma, D.~Shungin, H.~Mallick, M.~Schirmer, L.~H. Nguyen, R.~Kolde,
  E.~Franzosa, H.~Vlamakis, R.~Xavier, and C.~Huttenhower, ``Population
  structure discovery in meta-analyzed microbial communities and inflammatory
  bowel disease using mmuphin,'' {\em Genome Biology}, vol.~23, no.~1,
  pp.~1--31, 2022.

\bibitem{duvallet2017meta}
C.~Duvallet, S.~M. Gibbons, T.~Gurry, R.~A. Irizarry, and E.~J. Alm,
  ``Meta-analysis of gut microbiome studies identifies disease-specific and
  shared responses,'' {\em Nature communications}, vol.~8, no.~1, p.~1784,
  2017.

\bibitem{chong2020meta}
J.~Chong, P.~Liu, G.~Zhou, and J.~Xia, ``Using microbiomeanalyst for
  comprehensive statistical, functional, and meta-analysis of microbiome
  data,'' {\em Nature protocols}, vol.~15, no.~3, pp.~799--821, 2020.

\bibitem{Morton_meta_asd}
J.~T. Morton, D.-M. Jin, R.~H. Mills, Y.~Shao, G.~Rahman, D.~McDonald, Q.~Zhu,
  M.~Balaban, Y.~Jiang, K.~Cantrell, {\em et~al.}, ``Multi-level analysis of
  the gut--brain axis shows autism spectrum disorder-associated molecular and
  microbial profiles,'' {\em Nature Neuroscience}, pp.~1--10, 2023.

\end{thebibliography}

\end{document}


\maketitle

\hypertarget{introduction}{%
\section{0. Introduction}\label{introduction}}

Here, we will demonstrate four strategies to enrich microbiome-gut-brain
axis experiments. In this document we expand on the demonstration in the
supplementary files of the companion piece to this manuscript. We
strongly recommend readers go through that analysis before this
document, for the sake of continuity and clarity.

For this demonstration, we have adapted some shotgun metagenomic data
from the \texttt{curatedMetagenomicData} library in R. We're looking at
a human cohort starring in the \emph{Metagenome-wide association of gut
microbiome features for schizophrenia} study (DOI:
10.1038/s41467-020-15457-9). After downloading the data it was
simplified by summing together all strains by genus. This will make it
easier to analyse without access to a server. For the set of operations
used to pre-process, please see section
\href{https://github.com/thomazbastiaanssen/Tjazi/blob/master/guidebook_sup/part2/README_part2.md\#gathering-and-preparing-our-data-1}{Download
and pre-process microbiome data} in this document. Briefly, in this data
set, we have WGS data from faecal samples from both patients with
schizophrenia and healthy volunteers, which will be referred to as
``healthy'' in the Legends. This data has been included in the
\texttt{Tjazi} library on github for easy access purposes. All R code
used to transform, wrangle (reorganise) and plot the data is also shown
below as to hopefully provide a toolkit for aspiring and veteran
bioinformaticians alike.

\hypertarget{code-chunk-load-our-libraries}{%
\subsubsection{Code chunk: Load our
libraries}\label{code-chunk-load-our-libraries}}

\begin{Shaded}
\begin{Highlighting}[]
\CommentTok{\#Statistical tools        Primarily the CLR transformation.}
\FunctionTok{library}\NormalTok{(Tjazi)            }\CommentTok{\#devtools::install\_github("thomazbastiaanssen/Tjazi")}

\CommentTok{\#Data Wrangling}
\FunctionTok{library}\NormalTok{(tidyverse)        }\CommentTok{\#install.packages("tidyverse")}
\FunctionTok{library}\NormalTok{(knitr)            }\CommentTok{\#install.packages("knitr")}

\CommentTok{\#Plotting}
\FunctionTok{library}\NormalTok{(ggplot2)          }\CommentTok{\#install.packages("ggplot2")}
\FunctionTok{library}\NormalTok{(ggforce)          }\CommentTok{\#install.packages("ggforce")}
\FunctionTok{library}\NormalTok{(patchwork)        }\CommentTok{\#install.packages("patchwork")}

\CommentTok{\#Section specific packages}
\FunctionTok{library}\NormalTok{(mediation)        }\CommentTok{\#install.packages("mediation")}
\FunctionTok{library}\NormalTok{(anansi)           }\CommentTok{\#devtools::install\_github("thomazbastiaanssen/anansi")}
\FunctionTok{library}\NormalTok{(propr)            }\CommentTok{\#devtools::install\_github("tpq/propr")}
\FunctionTok{library}\NormalTok{(volatility)       }\CommentTok{\#devtools::install\_github("thomazbastiaanssen/volatility")}


\CommentTok{\#Load prepared data from the schizophrenia study stored in the Tjazi library}
\FunctionTok{data}\NormalTok{(guidebook\_data)}
\end{Highlighting}
\end{Shaded}

\newpage

\hypertarget{mediation-analysis}{%
\section{1. Mediation analysis}\label{mediation-analysis}}

Mediation analysis is a statistical tool that can help us answer causal
questions. Mediation analysis can be used to address core questions in
the microbiome-gut-brain axis field. One common and important example is
as follows:

\begin{itemize}
\tightlist
\item
  We know diet affects mental health.
\item
  We also know diet affects the microbiome.
\item
  We have observed associations between microbial taxa and mental
  health.
\end{itemize}

But how do we make sure that the association between the microbiome and
mental health isn't due to the fact that they share a common `driver',
namely diet, rendering our observed associations between the microbiome
and the brain merely spurious? Conversely, mediation analysis could help
us find to which degree the underlying mechanism of the effect of diet
on mental health is due to the effect of diet on the microbiome.

Here, we will demonstrate how one could go about performing mediation
analysis to ask and answer this question. We will use the 2020
schizophrenia data set to demonstrate this point. It should be noted
here that in reality, one would need strong biological mechanistic
reasons to perform a mediation analysis in order to make the `causal'
part in causal mediation analysis carry any meaning. Here we are just
performing the analysis for demonstrative purposes.

\hypertarget{gathering-and-preparing-our-data}{%
\subsection{Gathering and preparing our
data}\label{gathering-and-preparing-our-data}}

First, Let's load the schizophrenia data set into our environment. We'll
also quickly clean and CLR-transform our genus-level count table.

\hypertarget{code-chunk-preparing-microbiome-data}{%
\subsubsection{Code chunk: Preparing microbiome
data}\label{code-chunk-preparing-microbiome-data}}

\begin{Shaded}
\begin{Highlighting}[]
\CommentTok{\#Set a seed for the purposes of reproducibility in this document.}
\FunctionTok{set.seed}\NormalTok{(}\DecValTok{1}\NormalTok{)}

\CommentTok{\#Load the mediation library and the relevant files for demo}
\FunctionTok{library}\NormalTok{(mediation)}
\FunctionTok{data}\NormalTok{(guidebook\_data)}
\NormalTok{counts   }\OtherTok{\textless{}{-}}\NormalTok{ counts ; metadata }\OtherTok{\textless{}{-}}\NormalTok{ metadata ; diet }\OtherTok{\textless{}{-}}\NormalTok{ diet}

\CommentTok{\#Repeat the cleaning and transformation steps from part 1}
\NormalTok{metadata}\SpecialCharTok{$}\NormalTok{master\_ID }\OtherTok{\textless{}{-}} \FunctionTok{gsub}\NormalTok{(metadata}\SpecialCharTok{$}\NormalTok{master\_ID, }\AttributeTok{pattern =} \StringTok{"{-}"}\NormalTok{, }\AttributeTok{replacement =} \StringTok{"."}\NormalTok{)}

\NormalTok{counts  }\OtherTok{\textless{}{-}}\NormalTok{ counts[,metadata}\SpecialCharTok{$}\NormalTok{master\_ID]}

\CommentTok{\#Fork off your count data so that you always have an untouched version handy.}
\NormalTok{genus   }\OtherTok{\textless{}{-}}\NormalTok{ counts}

\CommentTok{\#make sure our count data is all numbers}
\NormalTok{genus   }\OtherTok{\textless{}{-}} \FunctionTok{apply}\NormalTok{(genus,}\FunctionTok{c}\NormalTok{(}\DecValTok{1}\NormalTok{,}\DecValTok{2}\NormalTok{),}\ControlFlowTok{function}\NormalTok{(x) }\FunctionTok{as.numeric}\NormalTok{(}\FunctionTok{as.character}\NormalTok{(x)))}

\CommentTok{\#Remove features with prevalence \textless{} 10\% in two steps:}
\CommentTok{\#First, determine how often every feature is absent in a sample}
\NormalTok{n\_zeroes }\OtherTok{\textless{}{-}} \FunctionTok{rowSums}\NormalTok{(genus }\SpecialCharTok{==} \DecValTok{0}\NormalTok{)}

\CommentTok{\#Then, remove features that are absent in more than your threshold (90\% in this case).}
\NormalTok{genus    }\OtherTok{\textless{}{-}}\NormalTok{ genus[n\_zeroes }\SpecialCharTok{\textless{}=} \FunctionTok{round}\NormalTok{(}\FunctionTok{ncol}\NormalTok{(genus) }\SpecialCharTok{*} \FloatTok{0.90}\NormalTok{),]}

\CommentTok{\#Perform a CLR transformation}
\NormalTok{genus.exp }\OtherTok{\textless{}{-}} \FunctionTok{clr\_c}\NormalTok{(genus)}
\end{Highlighting}
\end{Shaded}

\hypertarget{preparing-the-dietary-data}{%
\subsection{Preparing the dietary
data}\label{preparing-the-dietary-data}}

Furthermore, we will need to prepare the dietary data. For this study,
we have access to a table of five columns denoting how frequently our
participants eat food groups. It should be noted that we are only using
this data for demonstration purposes. In reality, the resolution of the
dietary information provided with this study would not be sufficient to
make any type of causal claim for several reasons. That said, let's
convert the dietary frequency categories into numbers and then perform a
principal component analysis.

\hypertarget{code-chunk-preparing-dietary-data}{%
\subsubsection{Code chunk: Preparing dietary
data}\label{code-chunk-preparing-dietary-data}}

\begin{Shaded}
\begin{Highlighting}[]
\NormalTok{diet.pca }\OtherTok{=}\NormalTok{ diet }\SpecialCharTok{\%\textgreater{}\%} 
  
\CommentTok{\#Replace text with numbers}
  \FunctionTok{mutate}\NormalTok{(}\FunctionTok{across}\NormalTok{(}\FunctionTok{everything}\NormalTok{(), }\SpecialCharTok{\textasciitilde{}}\FunctionTok{str\_replace}\NormalTok{( ., }\StringTok{"hardly"}\NormalTok{,                      }\StringTok{"1"}\NormalTok{ )),}
         \FunctionTok{across}\NormalTok{(}\FunctionTok{everything}\NormalTok{(), }\SpecialCharTok{\textasciitilde{}}\FunctionTok{str\_replace}\NormalTok{( ., }\StringTok{"often"}\NormalTok{,                       }\StringTok{"2"}\NormalTok{ )),}
         \FunctionTok{across}\NormalTok{(}\FunctionTok{everything}\NormalTok{(), }\SpecialCharTok{\textasciitilde{}}\FunctionTok{str\_replace}\NormalTok{( ., }\StringTok{"twice or three times a week"}\NormalTok{, }\StringTok{"3"}\NormalTok{ )),}
         \FunctionTok{across}\NormalTok{(}\FunctionTok{everything}\NormalTok{(), }\SpecialCharTok{\textasciitilde{}}\FunctionTok{str\_replace}\NormalTok{( ., }\StringTok{"every day"}\NormalTok{,                   }\StringTok{"4"}\NormalTok{ )), }
         \FunctionTok{across}\NormalTok{(}\FunctionTok{everything}\NormalTok{(), as.numeric)) }\SpecialCharTok{\%\textgreater{}\%} 

\CommentTok{\#Center and rescale data, then perform a principal component analysis}
  \FunctionTok{scale}\NormalTok{() }\SpecialCharTok{\%\textgreater{}\%}   
  \FunctionTok{prcomp}\NormalTok{()       }

\CommentTok{\#Let\textquotesingle{}s check out the loadings of the first principal component using a histogram }
\NormalTok{PC1 }\OtherTok{=}\NormalTok{ diet.pca}\SpecialCharTok{$}\NormalTok{x[,}\DecValTok{1}\NormalTok{]}
\FunctionTok{hist}\NormalTok{(PC1)}
\end{Highlighting}
\end{Shaded}

\includegraphics{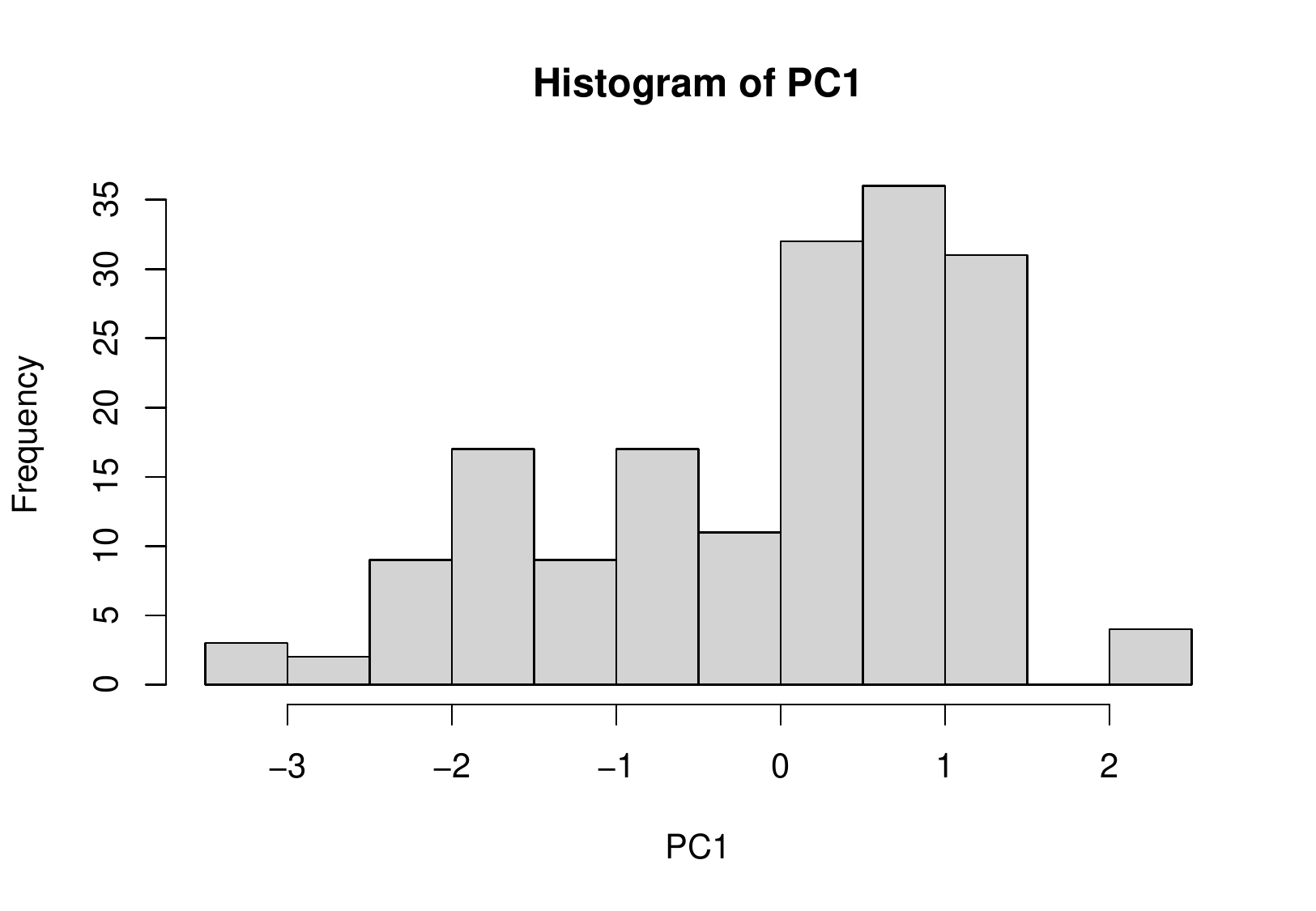}

It's always a good idea to visually inspect your data. From this
histogram, it looks like there may be two roughly normal populations in
our first principal component. We'll take note of this and proceed with
our preparations.

\hypertarget{prepare-data-for-mediation-analysis}{%
\subsection{Prepare data for mediation
analysis}\label{prepare-data-for-mediation-analysis}}

We'll take two of the bacteria, \emph{Cronobacter} and
\emph{Gordonibacter}, that were found to be statistically associated
with schizophrenia in our initial analysis. We'll also take the first
component of the dietary intake PCA we generated above.

\hypertarget{code-chunk-preparing-for-mediation-analysis}{%
\subsubsection{Code chunk: Preparing for mediation
analysis}\label{code-chunk-preparing-for-mediation-analysis}}

\begin{Shaded}
\begin{Highlighting}[]
\NormalTok{df\_mediation }\OtherTok{\textless{}{-}} \FunctionTok{data.frame}\NormalTok{(}
  \AttributeTok{Cronobacter    =} \FunctionTok{unlist}\NormalTok{(genus.exp[}\StringTok{"Enterobacteriaceae\_Cronobacter"}\NormalTok{,]),}
  \AttributeTok{Gordonibacter  =} \FunctionTok{unlist}\NormalTok{(genus.exp[}\StringTok{"Eggerthellaceae\_Gordonibacter"}\NormalTok{,]),}
  \AttributeTok{diet\_PC1       =}\NormalTok{ PC1, }
  \AttributeTok{phenotype      =}\NormalTok{ metadata}\SpecialCharTok{$}\NormalTok{Group }\SpecialCharTok{==} \StringTok{"schizophrenia"}\NormalTok{)}
\end{Highlighting}
\end{Shaded}

\newpage

\hypertarget{fit-models}{%
\subsection{Fit models}\label{fit-models}}

For mediation analysis, we need to consider a few models. Microbes will
be playing the role of (potential) mediator here.

\begin{itemize}
\item
  First, we estimate the effect of diet on our phenotype:
  \(phenotype \sim diet\). Does diet explain phenotype? If yes, we can
  proceed.
\item
  Second, we estimate the effect of the microbe on our phenotype of
  interest: \(phenotype \sim microbe\). Does the microbe also explain
  phenotype? If also yes, great, we have a potential mediation on our
  hands.
\item
  Third, we estimate the effect of diet on our microbe of interest:
  \(microbe \sim diet\). Does diet explain the abundance of our microbe?
  If yes, we can proceed. Now things are getting interesting as we have
  the scenario we laid out earlier.
\item
  Fourth, we estimate the \emph{joint} effect of diet and the microbe on
  our phenotype of interest: \(phenotype \sim diet + microbe\). Does
  diet now explain phenotype worse in the presence of the microbe? If
  so, we have a potential mediation on our hands. In the case that diet
  no longer explains phenotype at all, we may be dealing with a full
  mediation. in the case of a reduction of explanatory potential, we
  rather speak of a potential partial mediation.
\end{itemize}

To check for a mediation effect, we use the \texttt{mediation} package.

In our case, the phenotype (schizophrenia diagnosis) is a binary
outcome. Because of that, we'll use a logistic regression model rather
than a `regular' linear model whenever we're trying to explain
phenotype. We'll use a link function for this.

Let's give it a shot with our two bacteria.

\hypertarget{code-chunk-fitting-statistical-models}{%
\subsubsection{Code chunk: Fitting statistical
models}\label{code-chunk-fitting-statistical-models}}

\begin{Shaded}
\begin{Highlighting}[]
\CommentTok{\#Cronobacter}
\CommentTok{\#Does diet explain phenotype?}
\NormalTok{diet.fit1  }\OtherTok{\textless{}{-}} \FunctionTok{glm}\NormalTok{(phenotype }\SpecialCharTok{\textasciitilde{}}\NormalTok{ diet\_PC1, }
                  \AttributeTok{family =} \FunctionTok{binomial}\NormalTok{(}\StringTok{"logit"}\NormalTok{), }\AttributeTok{data =}\NormalTok{ df\_mediation)}

\CommentTok{\#Does diet explain Cronobacter?}
\NormalTok{diba.fit1  }\OtherTok{\textless{}{-}}  \FunctionTok{lm}\NormalTok{(Cronobacter }\SpecialCharTok{\textasciitilde{}}\NormalTok{ diet\_PC1, }\AttributeTok{data =}\NormalTok{ df\_mediation)}

\CommentTok{\#Does Gordonibacter explain phenotype?}
\NormalTok{bact.fit1  }\OtherTok{\textless{}{-}}  \FunctionTok{glm}\NormalTok{(phenotype }\SpecialCharTok{\textasciitilde{}}\NormalTok{ Cronobacter, }
                    \AttributeTok{family =} \FunctionTok{binomial}\NormalTok{(}\StringTok{"logit"}\NormalTok{), }\AttributeTok{data =}\NormalTok{ df\_mediation)}

\CommentTok{\#Does diet explain phenotype on the presence of Cronobacter?}
\NormalTok{both.fit1  }\OtherTok{\textless{}{-}} \FunctionTok{glm}\NormalTok{(phenotype }\SpecialCharTok{\textasciitilde{}}\NormalTok{ diet\_PC1 }\SpecialCharTok{+}\NormalTok{ Cronobacter, }
                  \AttributeTok{family =} \FunctionTok{binomial}\NormalTok{(}\StringTok{"logit"}\NormalTok{), }\AttributeTok{data =}\NormalTok{ df\_mediation)}

\CommentTok{\#Is there a mediation effect here?}
\NormalTok{crono }\OtherTok{=} \FunctionTok{mediate}\NormalTok{(diba.fit1, both.fit1, }
                \AttributeTok{treat =} \StringTok{\textquotesingle{}diet\_PC1\textquotesingle{}}\NormalTok{, }\AttributeTok{mediator =} \StringTok{\textquotesingle{}Cronobacter\textquotesingle{}}\NormalTok{, }\AttributeTok{boot =}\NormalTok{ T)}
\end{Highlighting}
\end{Shaded}

\begin{verbatim}
## Running nonparametric bootstrap
\end{verbatim}

\begin{Shaded}
\begin{Highlighting}[]
\CommentTok{\#Gordonibacter}
\CommentTok{\#Does diet explain phenotype?}
\NormalTok{diet.fit2  }\OtherTok{\textless{}{-}} \FunctionTok{glm}\NormalTok{(phenotype }\SpecialCharTok{\textasciitilde{}}\NormalTok{ diet\_PC1, }
                  \AttributeTok{family =} \FunctionTok{binomial}\NormalTok{(}\StringTok{"logit"}\NormalTok{), }\AttributeTok{data =}\NormalTok{ df\_mediation)}

\CommentTok{\#Does diet explain Gordonibacter?}
\NormalTok{diba.fit2  }\OtherTok{\textless{}{-}}  \FunctionTok{lm}\NormalTok{(Gordonibacter }\SpecialCharTok{\textasciitilde{}}\NormalTok{ diet\_PC1, }\AttributeTok{data =}\NormalTok{ df\_mediation)}

\CommentTok{\#Does Gordonibacter explain phenotype?}
\NormalTok{bact.fit2  }\OtherTok{\textless{}{-}}  \FunctionTok{glm}\NormalTok{(phenotype }\SpecialCharTok{\textasciitilde{}}\NormalTok{ Gordonibacter, }
                    \AttributeTok{family =} \FunctionTok{binomial}\NormalTok{(}\StringTok{"logit"}\NormalTok{), }\AttributeTok{data =}\NormalTok{ df\_mediation)}

\CommentTok{\#Does diet explain phenotype on the presence of Gordonibacter?}
\NormalTok{both.fit2  }\OtherTok{\textless{}{-}} \FunctionTok{glm}\NormalTok{(phenotype }\SpecialCharTok{\textasciitilde{}}\NormalTok{ diet\_PC1 }\SpecialCharTok{+}\NormalTok{ Gordonibacter, }
                  \AttributeTok{family =} \FunctionTok{binomial}\NormalTok{(}\StringTok{"logit"}\NormalTok{), }\AttributeTok{data =}\NormalTok{ df\_mediation)}

\CommentTok{\#Is there a mediation effect here?}
\NormalTok{gordo }\OtherTok{=} \FunctionTok{mediate}\NormalTok{(diba.fit2, both.fit2, }
                \AttributeTok{treat =} \StringTok{\textquotesingle{}diet\_PC1\textquotesingle{}}\NormalTok{, }\AttributeTok{mediator =} \StringTok{\textquotesingle{}Gordonibacter\textquotesingle{}}\NormalTok{, }\AttributeTok{boot =}\NormalTok{ T)}
\end{Highlighting}
\end{Shaded}

\begin{verbatim}
## Running nonparametric bootstrap
\end{verbatim}

Notice that in the \texttt{mediate} function calls, we're essentially
estimating the explanatory potential of diet on our phenotype, in the
presence of our bacteria, in light of the fact that diet also explains
our bacterial abundance.

\newpage

\hypertarget{investigate-results}{%
\subsection{Investigate results}\label{investigate-results}}

Let's take a look at the model summaries. We've picked the first
bacterium, \emph{Cronobacter}, to display a statistically significant
mediation effect.

\hypertarget{code-chunk-cronobacter-results}{%
\subsubsection{Code chunk: Cronobacter
results}\label{code-chunk-cronobacter-results}}

\begin{Shaded}
\begin{Highlighting}[]
\CommentTok{\#Collect the relevant data to display in tables}
\NormalTok{res\_diet.fit1 }\OtherTok{\textless{}{-}} \FunctionTok{coefficients}\NormalTok{(}\FunctionTok{summary}\NormalTok{(diet.fit1))}
\NormalTok{res\_diba.fit1 }\OtherTok{\textless{}{-}} \FunctionTok{coefficients}\NormalTok{(}\FunctionTok{summary}\NormalTok{(diba.fit1))}
\NormalTok{res\_bact.fit1 }\OtherTok{\textless{}{-}} \FunctionTok{coefficients}\NormalTok{(}\FunctionTok{summary}\NormalTok{(bact.fit1))}
\NormalTok{res\_both.fit1 }\OtherTok{\textless{}{-}} \FunctionTok{coefficients}\NormalTok{(}\FunctionTok{summary}\NormalTok{(both.fit1))}
                      
\CommentTok{\#Plot the results in nice looking tables:}
\FunctionTok{kable}\NormalTok{(res\_diet.fit1, }\AttributeTok{digits =} \DecValTok{3}\NormalTok{, }
      \AttributeTok{caption =} \StringTok{"Diet significantly explains phenotype (Estimate of 0.241)."}\NormalTok{)}
\end{Highlighting}
\end{Shaded}

\begin{longtable}[]{@{}lrrrr@{}}
\caption{Diet significantly explains phenotype (Estimate of
0.241).}\tabularnewline
\toprule()
& Estimate & Std. Error & z value &
Pr(\textgreater\textbar z\textbar) \\
\midrule()
\endfirsthead
\toprule()
& Estimate & Std. Error & z value &
Pr(\textgreater\textbar z\textbar) \\
\midrule()
\endhead
(Intercept) & 0.106 & 0.155 & 0.687 & 0.492 \\
diet\_PC1 & 0.241 & 0.122 & 1.972 & 0.049 \\
\bottomrule()
\end{longtable}

\begin{Shaded}
\begin{Highlighting}[]
\FunctionTok{kable}\NormalTok{(res\_diba.fit1, }\AttributeTok{digits =} \DecValTok{3}\NormalTok{, }
      \AttributeTok{caption =} \StringTok{"Diet significantly explains Cronobacter."}\NormalTok{)}
\end{Highlighting}
\end{Shaded}

\begin{longtable}[]{@{}lrrrr@{}}
\caption{Diet significantly explains Cronobacter.}\tabularnewline
\toprule()
& Estimate & Std. Error & t value &
Pr(\textgreater\textbar t\textbar) \\
\midrule()
\endfirsthead
\toprule()
& Estimate & Std. Error & t value &
Pr(\textgreater\textbar t\textbar) \\
\midrule()
\endhead
(Intercept) & -2.461 & 0.114 & -21.676 & 0.000 \\
diet\_PC1 & 0.182 & 0.088 & 2.061 & 0.041 \\
\bottomrule()
\end{longtable}

\begin{Shaded}
\begin{Highlighting}[]
\FunctionTok{kable}\NormalTok{(res\_bact.fit1, }\AttributeTok{digits =} \DecValTok{3}\NormalTok{, }
      \AttributeTok{caption =} \StringTok{"Cronobacter significantly explains phenotype. This comes as }
\StringTok{      no surprise as saw this in our initial differential abundance analysis."}\NormalTok{)}
\end{Highlighting}
\end{Shaded}

\begin{longtable}[]{@{}lrrrr@{}}
\caption{Cronobacter significantly explains phenotype. This comes as no
surprise as saw this in our initial differential abundance
analysis.}\tabularnewline
\toprule()
& Estimate & Std. Error & z value &
Pr(\textgreater\textbar z\textbar) \\
\midrule()
\endfirsthead
\toprule()
& Estimate & Std. Error & z value &
Pr(\textgreater\textbar z\textbar) \\
\midrule()
\endhead
(Intercept) & 0.882 & 0.332 & 2.655 & 0.008 \\
Cronobacter & 0.310 & 0.114 & 2.722 & 0.006 \\
\bottomrule()
\end{longtable}

\begin{Shaded}
\begin{Highlighting}[]
\FunctionTok{kable}\NormalTok{(res\_both.fit1, }\AttributeTok{digits =} \DecValTok{3}\NormalTok{, }
      \AttributeTok{caption =} \StringTok{"In the presence of Cronobacter, diet significantly }
\StringTok{      explains phenotype less well (Estimate of 0.199)."}\NormalTok{)}
\end{Highlighting}
\end{Shaded}

\begin{longtable}[]{@{}lrrrr@{}}
\caption{In the presence of Cronobacter, diet significantly explains
phenotype less well (Estimate of 0.199).}\tabularnewline
\toprule()
& Estimate & Std. Error & z value &
Pr(\textgreater\textbar z\textbar) \\
\midrule()
\endfirsthead
\toprule()
& Estimate & Std. Error & z value &
Pr(\textgreater\textbar z\textbar) \\
\midrule()
\endhead
(Intercept) & 0.831 & 0.337 & 2.467 & 0.014 \\
diet\_PC1 & 0.199 & 0.124 & 1.601 & 0.109 \\
Cronobacter & 0.288 & 0.115 & 2.496 & 0.013 \\
\bottomrule()
\end{longtable}

Looks like we may have a mediation effect here, so let's check out the
results of our mediation analysis: ACME stands for Average Causal
Mediation Effect, whereas ADE stands for Average Direct Effect.

\hypertarget{code-chunk-cronobacter-mediation-figure}{%
\subsubsection{Code chunk: Cronobacter mediation
figure}\label{code-chunk-cronobacter-mediation-figure}}

\begin{Shaded}
\begin{Highlighting}[]
\FunctionTok{summary}\NormalTok{(crono)}
\end{Highlighting}
\end{Shaded}

\begin{verbatim}
## 
## Causal Mediation Analysis 
## 
## Nonparametric Bootstrap Confidence Intervals with the Percentile Method
## 
##                           Estimate 95% CI Lower 95% CI Upper p-value  
## ACME (control)            0.012500     0.000837         0.03   0.030 *
## ACME (treated)            0.012209     0.000834         0.03   0.030 *
## ADE (control)             0.047085    -0.008466         0.10   0.094 .
## ADE (treated)             0.046795    -0.008487         0.10   0.094 .
## Total Effect              0.059294     0.006317         0.12   0.030 *
## Prop. Mediated (control)  0.210805    -0.011539         1.18   0.060 .
## Prop. Mediated (treated)  0.205907    -0.011134         1.19   0.060 .
## ACME (average)            0.012354     0.000836         0.03   0.030 *
## ADE (average)             0.046940    -0.008476         0.10   0.094 .
## Prop. Mediated (average)  0.208356    -0.011337         1.19   0.060 .
## ---
## Signif. codes:  0 '***' 0.001 '**' 0.01 '*' 0.05 '.' 0.1 ' ' 1
## 
## Sample Size Used: 171 
## 
## 
## Simulations: 1000
\end{verbatim}

\begin{Shaded}
\begin{Highlighting}[]
\CommentTok{\#Let\textquotesingle{}s view the same information in a figure:}
\FunctionTok{plot}\NormalTok{(crono, }\AttributeTok{main =} \StringTok{"Cronobacter as a mediator of the effect of diet on schizophrenia"}\NormalTok{)}
\end{Highlighting}
\end{Shaded}

\includegraphics{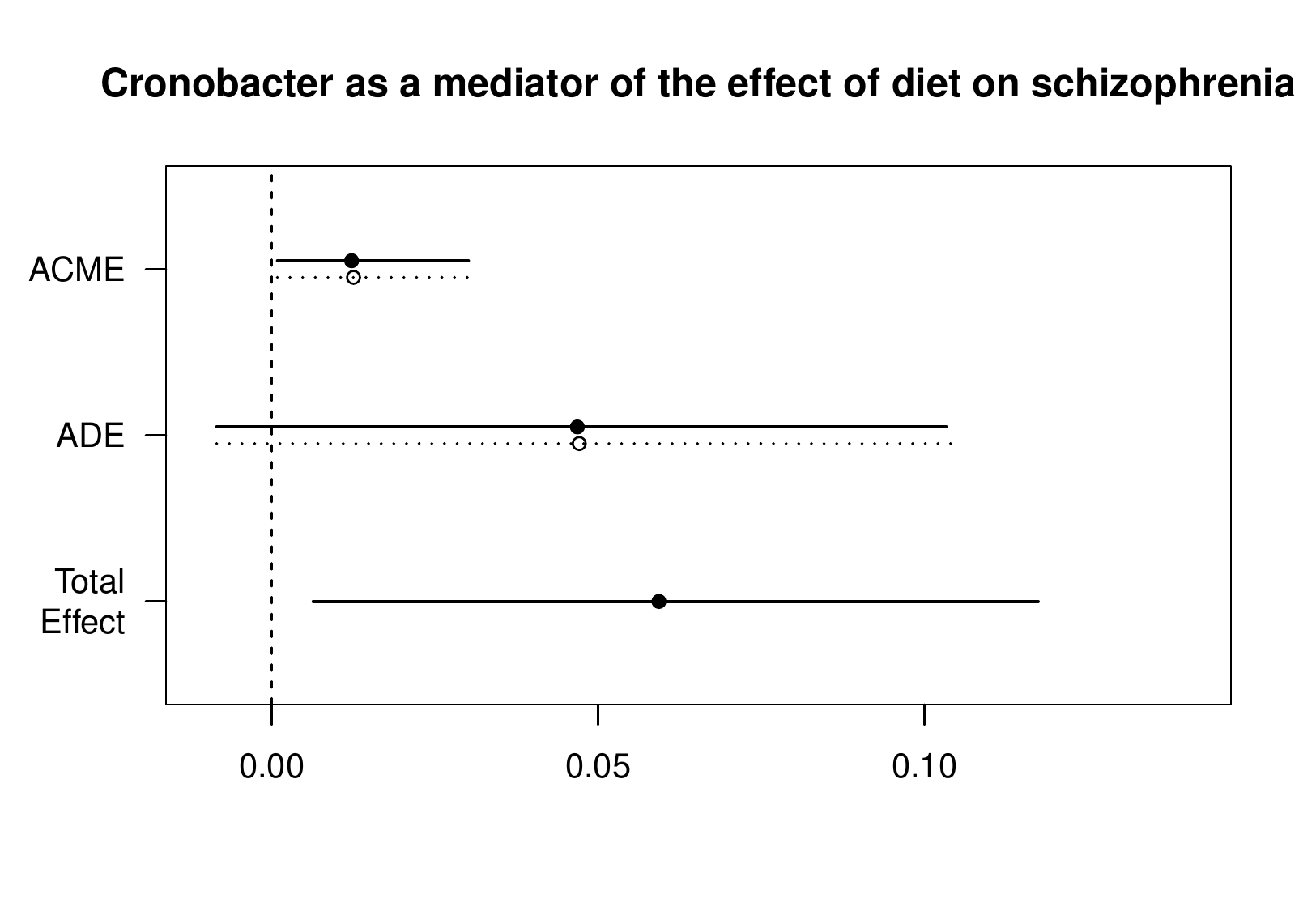}

Here, we see that our estimate of the Average Causal Mediation Effect
lies outside of zero. We have partial mediation. In other words, in this
example, it could be the case that part of the effect of diet on
schizophrenia happens because diet also influences \emph{Cronobacter},
which in turn influences schizophrenia.

\newpage

Now, let's take a look at a negative example with \emph{Gordonibacter}.

\hypertarget{code-chunk-gordonibacter-results}{%
\subsubsection{Code chunk: Gordonibacter
results}\label{code-chunk-gordonibacter-results}}

\begin{Shaded}
\begin{Highlighting}[]
\CommentTok{\#Collect the relevant data to display in tables}
\NormalTok{res\_diet.fit2 }\OtherTok{\textless{}{-}} \FunctionTok{coefficients}\NormalTok{(}\FunctionTok{summary}\NormalTok{(diet.fit2))}
\NormalTok{res\_diba.fit2 }\OtherTok{\textless{}{-}} \FunctionTok{coefficients}\NormalTok{(}\FunctionTok{summary}\NormalTok{(diba.fit2))}
\NormalTok{res\_bact.fit2 }\OtherTok{\textless{}{-}} \FunctionTok{coefficients}\NormalTok{(}\FunctionTok{summary}\NormalTok{(bact.fit2))}
\NormalTok{res\_both.fit2 }\OtherTok{\textless{}{-}} \FunctionTok{coefficients}\NormalTok{(}\FunctionTok{summary}\NormalTok{(both.fit2))}
                      
\CommentTok{\#Plot the results in nice looking tables:}
\FunctionTok{kable}\NormalTok{(res\_diet.fit2, }\AttributeTok{digits =} \DecValTok{3}\NormalTok{, }
      \AttributeTok{caption =} \StringTok{"Diet significantly explains phenotype (Estimate of 0.241)."}\NormalTok{)}
\end{Highlighting}
\end{Shaded}

\begin{longtable}[]{@{}lrrrr@{}}
\caption{Diet significantly explains phenotype (Estimate of
0.241).}\tabularnewline
\toprule()
& Estimate & Std. Error & z value &
Pr(\textgreater\textbar z\textbar) \\
\midrule()
\endfirsthead
\toprule()
& Estimate & Std. Error & z value &
Pr(\textgreater\textbar z\textbar) \\
\midrule()
\endhead
(Intercept) & 0.106 & 0.155 & 0.687 & 0.492 \\
diet\_PC1 & 0.241 & 0.122 & 1.972 & 0.049 \\
\bottomrule()
\end{longtable}

\begin{Shaded}
\begin{Highlighting}[]
\FunctionTok{kable}\NormalTok{(res\_diba.fit2, }\AttributeTok{digits =} \DecValTok{3}\NormalTok{, }
      \AttributeTok{caption =} \StringTok{"Diet does not significantly explain Gordonibacter"}\NormalTok{)}
\end{Highlighting}
\end{Shaded}

\begin{longtable}[]{@{}lrrrr@{}}
\caption{Diet does not significantly explain
Gordonibacter}\tabularnewline
\toprule()
& Estimate & Std. Error & t value &
Pr(\textgreater\textbar t\textbar) \\
\midrule()
\endfirsthead
\toprule()
& Estimate & Std. Error & t value &
Pr(\textgreater\textbar t\textbar) \\
\midrule()
\endhead
(Intercept) & -1.734 & 0.114 & -15.175 & 0.000 \\
diet\_PC1 & -0.037 & 0.089 & -0.410 & 0.682 \\
\bottomrule()
\end{longtable}

\begin{Shaded}
\begin{Highlighting}[]
\FunctionTok{kable}\NormalTok{(res\_bact.fit2, }\AttributeTok{digits =} \DecValTok{3}\NormalTok{, }
      \AttributeTok{caption =} \StringTok{"Gordonibacter significantly explains phenotype. This comes as }
\StringTok{      no surprise as saw this in our initial differential abundance analysis."}\NormalTok{)}
\end{Highlighting}
\end{Shaded}

\begin{longtable}[]{@{}lrrrr@{}}
\caption{Gordonibacter significantly explains phenotype. This comes as
no surprise as saw this in our initial differential abundance
analysis.}\tabularnewline
\toprule()
& Estimate & Std. Error & z value &
Pr(\textgreater\textbar z\textbar) \\
\midrule()
\endfirsthead
\toprule()
& Estimate & Std. Error & z value &
Pr(\textgreater\textbar z\textbar) \\
\midrule()
\endhead
(Intercept) & 0.607 & 0.252 & 2.409 & 0.016 \\
Gordonibacter & 0.285 & 0.110 & 2.592 & 0.010 \\
\bottomrule()
\end{longtable}

\begin{Shaded}
\begin{Highlighting}[]
\FunctionTok{kable}\NormalTok{(res\_both.fit2, }\AttributeTok{digits =} \DecValTok{3}\NormalTok{, }
      \AttributeTok{caption =} \StringTok{"In the presence of Gordonibacter, diet explains phenotype even better}
\StringTok{      (Estimate of 0.263)."}\NormalTok{)}
\end{Highlighting}
\end{Shaded}

\begin{longtable}[]{@{}lrrrr@{}}
\caption{In the presence of Gordonibacter, diet explains phenotype even
better (Estimate of 0.263).}\tabularnewline
\toprule()
& Estimate & Std. Error & z value &
Pr(\textgreater\textbar z\textbar) \\
\midrule()
\endfirsthead
\toprule()
& Estimate & Std. Error & z value &
Pr(\textgreater\textbar z\textbar) \\
\midrule()
\endhead
(Intercept) & 0.633 & 0.255 & 2.484 & 0.013 \\
diet\_PC1 & 0.263 & 0.125 & 2.099 & 0.036 \\
Gordonibacter & 0.299 & 0.111 & 2.686 & 0.007 \\
\bottomrule()
\end{longtable}

Looks like we may have a mediation effect here, so let's check out the
results of our mediation analysis: Again, ACME stands for Average Causal
Mediation Effect, whereas ADE stands for Average Direct Effect.

\hypertarget{code-chunk-gordonibacter-mediation-figure}{%
\subsubsection{Code chunk: Gordonibacter mediation
figure}\label{code-chunk-gordonibacter-mediation-figure}}

\begin{Shaded}
\begin{Highlighting}[]
\FunctionTok{summary}\NormalTok{(gordo)}
\end{Highlighting}
\end{Shaded}

\begin{verbatim}
## 
## Causal Mediation Analysis 
## 
## Nonparametric Bootstrap Confidence Intervals with the Percentile Method
## 
##                          Estimate 95% CI Lower 95% CI Upper p-value  
## ACME (control)           -0.00260     -0.01755         0.01   0.720  
## ACME (treated)           -0.00253     -0.01708         0.01   0.720  
## ADE (control)             0.06191      0.00675         0.12   0.038 *
## ADE (treated)             0.06198      0.00675         0.12   0.038 *
## Total Effect              0.05938      0.00130         0.12   0.050 *
## Prop. Mediated (control) -0.04374     -0.67937         0.36   0.758  
## Prop. Mediated (treated) -0.04266     -0.66939         0.36   0.758  
## ACME (average)           -0.00257     -0.01749         0.01   0.720  
## ADE (average)             0.06194      0.00675         0.12   0.038 *
## Prop. Mediated (average) -0.04320     -0.67536         0.36   0.758  
## ---
## Signif. codes:  0 '***' 0.001 '**' 0.01 '*' 0.05 '.' 0.1 ' ' 1
## 
## Sample Size Used: 171 
## 
## 
## Simulations: 1000
\end{verbatim}

\begin{Shaded}
\begin{Highlighting}[]
\CommentTok{\#Let\textquotesingle{}s view the same information in a figure:}
\FunctionTok{plot}\NormalTok{(gordo, }\AttributeTok{main =} \StringTok{"Gordonibacter as a mediator of the effect of diet on schizophrenia"}\NormalTok{)}
\end{Highlighting}
\end{Shaded}

\includegraphics{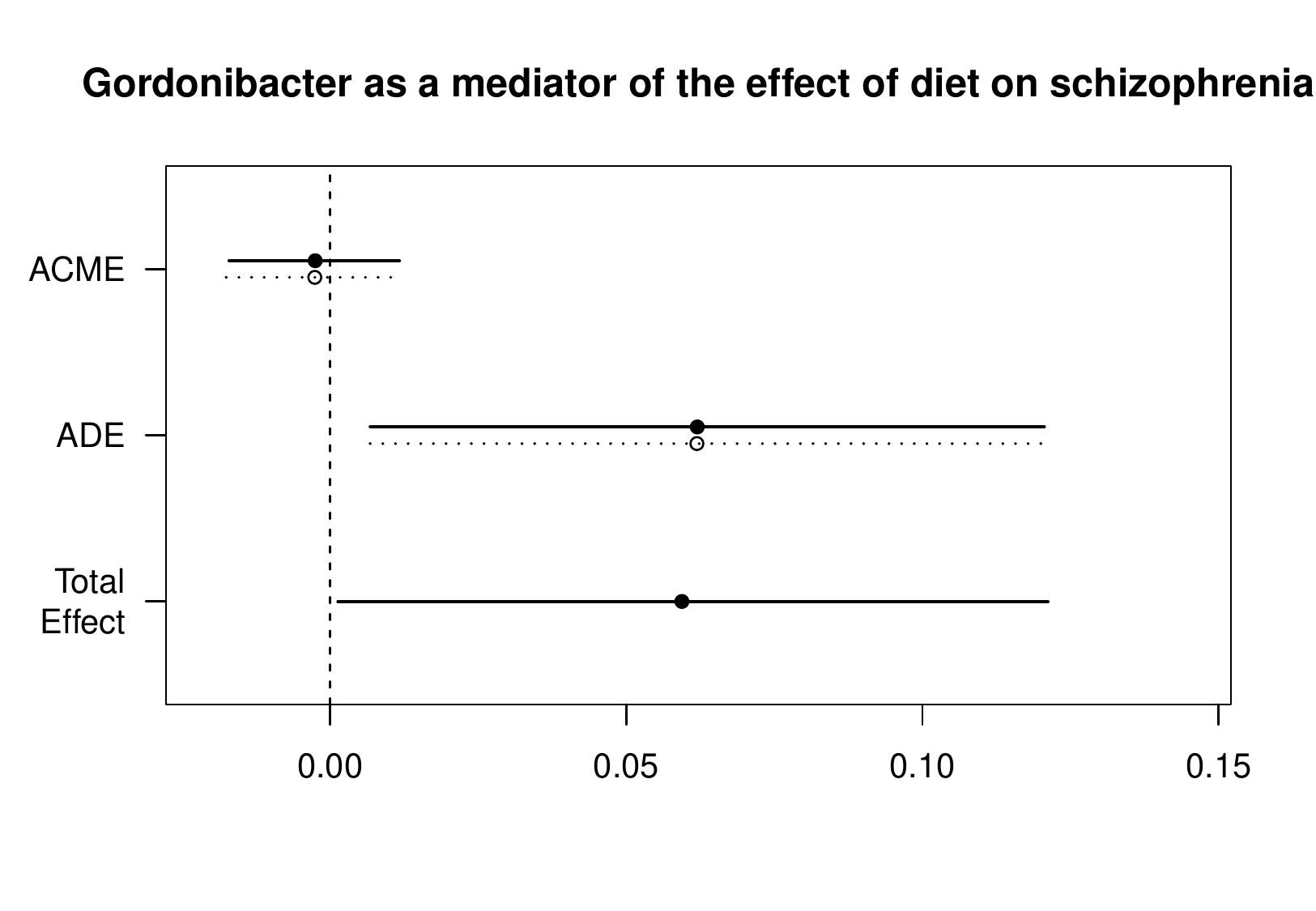}

Here, we see that our estimate of the Average Causal Mediation Effect
lies squarely on zero. No mediation. This makes sense, as diet couldn't
significantly explain the abundance of Gordonibacter.

\begin{center}\rule{0.5\linewidth}{0.5pt}\end{center}

\newpage

\hypertarget{multi-omics-integration}{%
\section{2. Multi-omics integration}\label{multi-omics-integration}}

Here, we will demonstrate how to integrate and subsequently analyse two
'omics data sets. For this, we will use the \texttt{anansi} framework
and package. Studies including both microbiome and metabolomics data are
becoming more common. Often, it would be helpful to integrate both data
sets in order to see if they corroborate each others patterns. All vs
all association is imprecise and likely to yield spurious associations.

The \texttt{anansi} package computes and compares the association
between the features of two 'omics data sets that are known to interact
based on a database such as KEGG. \texttt{anansi} takes a
knowledge-based approach to constrain association search space, only
considering metabolite-function interactions that have been recorded in
a pathway database.

We'll load a complementary training data set using
\texttt{data(FMT\_data)}. This loads a curated snippet from the data set
described in more detail here:
\url{https://doi.org/10.1038/s43587-021-00093-9}\\
A very early version of \texttt{anansi} was used to generate ``Extended
Data Fig. 7'' in that paper.

\hypertarget{code-chunk-load-libraries-and-data}{%
\subsubsection{Code chunk: Load libraries and
data}\label{code-chunk-load-libraries-and-data}}

\begin{Shaded}
\begin{Highlighting}[]
\CommentTok{\#install and load anansi}
\CommentTok{\#devtools::install\_github("thomazbastiaanssen/anansi")}
\FunctionTok{library}\NormalTok{(anansi)}

\CommentTok{\#load ggplot2 and ggforce to plot results}
\FunctionTok{library}\NormalTok{(ggplot2)}
\FunctionTok{library}\NormalTok{(ggforce)}

\CommentTok{\#load dictionary and complementary human{-}readable names for KEGG compounds and orthologues}
\FunctionTok{data}\NormalTok{(dictionary)}
\CommentTok{\#load example data + metadata from FMT Aging study}
\FunctionTok{data}\NormalTok{(FMT\_data)}
\end{Highlighting}
\end{Shaded}

\hypertarget{data-preparation}{%
\subsection{Data preparation}\label{data-preparation}}

The main \texttt{anansi} function expects data in the \texttt{anansiWeb}
format; Basically a list with exactly three tables: The first table,
\texttt{tableY}, should be a count table of metabolites. The second
table, \texttt{tableX}, should be a count table of functions. Both
tables should have columns as features and rows as samples.

The third table should be a binary adjacency matrix with the column
names of \texttt{tableY} as rows and the column names of \texttt{tableX}
as columns. Such an adjacency matrix is provided in the \texttt{anansi}
library and is referred to as a \texttt{dictionary} (because you use it
to look up which metabolites interact with which functions). We can load
it using \texttt{data(dictionary)}.

Though this example uses metabolites and functions, \texttt{anansi} is
able to handle any type of 'omics data, as long as there is a dictionary
available. Because of this, \texttt{anansi} uses the type-naive
nomenclature \texttt{tableY} and \texttt{tableX}. The Y and X refer to
the position these measurements will have in the linear modeling
framework:

\[Y \sim X \times {\text{covariates}}\] \newpage

\hypertarget{a-note-on-functional-microbiome-data}{%
\subsubsection{A note on functional microbiome
data}\label{a-note-on-functional-microbiome-data}}

Two common questions in the host-microbiome field are ``Who's there?''
and ``What are they doing?''.\\
Techniques like 16S sequencing and shotgun metagenomics sequencing are
most commonly used to answer the first question. The second question can
be a bit more tricky - often we'll need functional inference software to
address them. For 16S sequencing, algorithms like \texttt{PICRUSt2} and
\texttt{Piphillin} can be used to infer function. For shotgun
metagenomics, \texttt{HUMANn3} in the bioBakery suite or \texttt{woltka}
can be used.\\
All of these algorithms can produce functional count data in terms of
KEGG Orthologues (KOs). These tables can be directly plugged in to
\texttt{anansi}.

\hypertarget{code-chunk-prepare-data}{%
\subsubsection{Code chunk: Prepare data}\label{code-chunk-prepare-data}}

\begin{Shaded}
\begin{Highlighting}[]
\CommentTok{\#Clean and CLR{-}transform the KEGG orthologue table.}

\CommentTok{\#Only keep functions that are represented in the dictionary.}
\NormalTok{KOs     }\OtherTok{\textless{}{-}}\NormalTok{ FMT\_KOs[}\FunctionTok{row.names}\NormalTok{(FMT\_KOs) }\SpecialCharTok{\%in\%} \FunctionTok{sort}\NormalTok{(}\FunctionTok{unique}\NormalTok{(}\FunctionTok{unlist}\NormalTok{(anansi\_dic))),]}

\CommentTok{\#Cut the decimal part off.}
\NormalTok{KOs     }\OtherTok{\textless{}{-}} \FunctionTok{floor}\NormalTok{(KOs)}

\CommentTok{\#Ensure all entires are numbers.}
\NormalTok{KOs     }\OtherTok{\textless{}{-}} \FunctionTok{apply}\NormalTok{(KOs,}\FunctionTok{c}\NormalTok{(}\DecValTok{1}\NormalTok{,}\DecValTok{2}\NormalTok{),}\ControlFlowTok{function}\NormalTok{(x) }\FunctionTok{as.numeric}\NormalTok{(}\FunctionTok{as.character}\NormalTok{(x)))}

\CommentTok{\#Remove all features with \textless{} 10\% prevalence in the dataset.}
\NormalTok{KOs     }\OtherTok{\textless{}{-}}\NormalTok{ KOs[}\FunctionTok{apply}\NormalTok{(KOs }\SpecialCharTok{==} \DecValTok{0}\NormalTok{, }\DecValTok{1}\NormalTok{, sum) }\SpecialCharTok{\textless{}=}\NormalTok{ (}\FunctionTok{ncol}\NormalTok{(KOs) }\SpecialCharTok{*} \FloatTok{0.90}\NormalTok{), ] }

\CommentTok{\#Perform a centered log{-}ratio transformation on the functional count table.}
\NormalTok{KOs.exp }\OtherTok{\textless{}{-}} \FunctionTok{clr\_c}\NormalTok{(KOs)}

\CommentTok{\#anansi expects samples to be rows and features to be columns. }
\NormalTok{t1      }\OtherTok{\textless{}{-}} \FunctionTok{t}\NormalTok{(FMT\_metab)}
\NormalTok{t2      }\OtherTok{\textless{}{-}} \FunctionTok{t}\NormalTok{(KOs.exp)}
\end{Highlighting}
\end{Shaded}

\hypertarget{weave-a-web}{%
\subsection{Weave a web}\label{weave-a-web}}

The \texttt{weaveWebFromTables()} function can be used to parse the
tables that we prepared above into an \texttt{anansiWeb} object. The
\texttt{anansiWeb} format is a necessary input file for the main
\texttt{anansi} workflow.

\hypertarget{code-chunk-generate-web-object}{%
\subsubsection{Code chunk: Generate web
object}\label{code-chunk-generate-web-object}}

\begin{Shaded}
\begin{Highlighting}[]
\NormalTok{web }\OtherTok{\textless{}{-}} \FunctionTok{weaveWebFromTables}\NormalTok{(}\AttributeTok{tableY =}\NormalTok{ t1, }\AttributeTok{tableX =}\NormalTok{ t2, }\AttributeTok{dictionary =}\NormalTok{ anansi\_dic)}
\end{Highlighting}
\end{Shaded}

\begin{verbatim}
## [1] "Operating in interaction mode"
## [1] "3 were matched between table 1 and the columns of the adjacency matrix"
## [1] "50 were matched between table 2 and the rows of the adjacency matrix"
\end{verbatim}

\newpage

\hypertarget{run-anansi}{%
\subsection{Run anansi}\label{run-anansi}}

The main workspider in this package is called \texttt{anansi}.
Generally, you want to give it three arguments. First, there's
\texttt{web}, which is an \texttt{anansiWeb} object, such as the one we
generated in the above step. Second, there's \texttt{formula}, which
should be a formula. For instance, to assess differential associations
between treatments, we use the formula
\texttt{\textasciitilde{}Treatment}, provided we have a column with that
name in our \texttt{metadata} object, the Third argument.

\hypertarget{code-chunk-run-anansi}{%
\subsubsection{Code chunk: Run anansi}\label{code-chunk-run-anansi}}

\begin{Shaded}
\begin{Highlighting}[]
\NormalTok{anansi\_out }\OtherTok{\textless{}{-}} \FunctionTok{anansi}\NormalTok{(}\AttributeTok{web      =}\NormalTok{ web,          }\CommentTok{\#Generated above}
                     \AttributeTok{method   =} \StringTok{"pearson"}\NormalTok{,    }\CommentTok{\#Define the type of correlation used}
                     \AttributeTok{formula  =} \SpecialCharTok{\textasciitilde{}}\NormalTok{ Legend,     }\CommentTok{\#Compare associations between treatments}
                     \AttributeTok{metadata =}\NormalTok{ FMT\_metadata, }\CommentTok{\#With data referred to in formula as column}
                     \AttributeTok{adjust.method =} \StringTok{"BH"}\NormalTok{,    }\CommentTok{\#Apply the Benjamini{-}Hochberg procedure}
                     \AttributeTok{verbose  =}\NormalTok{ T             }\CommentTok{\#To let you know what\textquotesingle{}s happening}
\NormalTok{                     )}
\end{Highlighting}
\end{Shaded}

\begin{verbatim}
## [1] "Running annotation-based correlations"
## [1] "Running correlations for the following groups: All, Aged yFMT, Aged oFMT, Young yFMT"
## [1] "Fitting models for differential correlation testing"
## [1] "Model type:lm"
## [1] "Adjusting p-values using Benjamini & Hochberg's procedure."
## [1] "Using theoretical distribution."
\end{verbatim}

\hypertarget{spin-to-a-table}{%
\subsection{Spin to a table}\label{spin-to-a-table}}

\texttt{anansi} gives a complex nested \texttt{anansiYarn} object as an
output. Two functions exist that will wrangle your data to more friendly
formats for you. You can either use \texttt{spinToLong()} or
\texttt{spinToWide()}. They will give you long or wide format
data.frames, respectively. For general reporting, we recommend sticking
to the wide format as it's the most legible. You can also use the
\texttt{plot()} method on an \texttt{anansiYarn} object to gain some
insights in the state of your p, q, R and R\textsuperscript{2}
parameters.

\hypertarget{code-chunk-parse-anansi-results}{%
\subsubsection{Code chunk: Parse anansi
results}\label{code-chunk-parse-anansi-results}}

\begin{Shaded}
\begin{Highlighting}[]
\NormalTok{anansiLong }\OtherTok{\textless{}{-}} \FunctionTok{spinToLong}\NormalTok{(}\AttributeTok{anansi\_output =}\NormalTok{ anansi\_out, }\AttributeTok{translate =}\NormalTok{ T, }
                         \AttributeTok{Y\_translation =}\NormalTok{ anansi}\SpecialCharTok{::}\NormalTok{cpd\_translation, }
                         \AttributeTok{X\_translation =}\NormalTok{ anansi}\SpecialCharTok{::}\NormalTok{KO\_translation)  }
\CommentTok{\#Now it\textquotesingle{}s ready to be plugged into ggplot2, though let\textquotesingle{}s clean up a bit more. }

\CommentTok{\#Only consider interactions where the entire model fits well enough. }
\NormalTok{anansiLong }\OtherTok{\textless{}{-}}\NormalTok{ anansiLong[anansiLong}\SpecialCharTok{$}\NormalTok{model\_full\_q.values }\SpecialCharTok{\textless{}} \FloatTok{0.1}\NormalTok{,]}
\end{Highlighting}
\end{Shaded}

\newpage

\hypertarget{plot-the-results}{%
\subsection{Plot the results}\label{plot-the-results}}

The long format can be helpful to plug the data into \texttt{ggplot2}.
Here, we recreate part of the results from the FMT Aging study.

\hypertarget{code-chunk-plot-anansi-results}{%
\subsubsection{Code chunk: Plot anansi
results}\label{code-chunk-plot-anansi-results}}

\begin{Shaded}
\begin{Highlighting}[]
\FunctionTok{ggplot}\NormalTok{(}\AttributeTok{data =}\NormalTok{ anansiLong, }
       \FunctionTok{aes}\NormalTok{(}\AttributeTok{x      =}\NormalTok{ r.values, }
           \AttributeTok{y      =}\NormalTok{ feature\_X, }
           \AttributeTok{fill   =}\NormalTok{ type, }
           \AttributeTok{alpha  =}\NormalTok{ model\_disjointed\_Legend\_p.values }\SpecialCharTok{\textless{}} \FloatTok{0.05}\NormalTok{)) }\SpecialCharTok{+} 
  
  \CommentTok{\#Make a vertical dashed red line at x = 0}
  \FunctionTok{geom\_vline}\NormalTok{(}\AttributeTok{xintercept =} \DecValTok{0}\NormalTok{, }\AttributeTok{linetype =} \StringTok{"dashed"}\NormalTok{, }\AttributeTok{colour =} \StringTok{"red"}\NormalTok{)}\SpecialCharTok{+}
  
  \CommentTok{\#Points show  raw correlation coefficients}
  \FunctionTok{geom\_point}\NormalTok{(}\AttributeTok{shape =} \DecValTok{21}\NormalTok{, }\AttributeTok{size =} \DecValTok{3}\NormalTok{) }\SpecialCharTok{+} 
  
  \CommentTok{\#facet per compound}
\NormalTok{  ggforce}\SpecialCharTok{::}\FunctionTok{facet\_col}\NormalTok{(}\SpecialCharTok{\textasciitilde{}}\NormalTok{feature\_Y, }\AttributeTok{space =} \StringTok{"free"}\NormalTok{, }\AttributeTok{scales =} \StringTok{"free\_y"}\NormalTok{) }\SpecialCharTok{+} 
  
  \CommentTok{\#fix the scales, labels, theme and other layout}
  \FunctionTok{scale\_y\_discrete}\NormalTok{(}\AttributeTok{limits =}\NormalTok{ rev, }\AttributeTok{position =} \StringTok{"right"}\NormalTok{) }\SpecialCharTok{+}
  \FunctionTok{scale\_alpha\_manual}\NormalTok{(}\AttributeTok{values =} \FunctionTok{c}\NormalTok{(}\StringTok{"TRUE"} \OtherTok{=} \DecValTok{1}\NormalTok{, }
                                \StringTok{"FALSE"} \OtherTok{=} \DecValTok{1}\SpecialCharTok{/}\DecValTok{3}\NormalTok{), }\StringTok{"Disjointed association}\SpecialCharTok{\textbackslash{}n}\StringTok{p \textless{} 0.05"}\NormalTok{) }\SpecialCharTok{+}
  \FunctionTok{scale\_fill\_manual}\NormalTok{(}\AttributeTok{values =} \FunctionTok{c}\NormalTok{(}\StringTok{"Young yFMT"} \OtherTok{=} \StringTok{"\#2166ac"}\NormalTok{, }
                               \StringTok{"Aged oFMT"}  \OtherTok{=} \StringTok{"\#b2182b"}\NormalTok{, }
                               \StringTok{"Aged yFMT"}  \OtherTok{=} \StringTok{"\#ef8a62"}\NormalTok{, }
                               \StringTok{"All"}        \OtherTok{=} \StringTok{"gray"}\NormalTok{), }
                    \AttributeTok{breaks =} \FunctionTok{c}\NormalTok{(}\StringTok{"Young yFMT"}\NormalTok{, }\StringTok{"Aged oFMT"}\NormalTok{, }\StringTok{"Aged yFMT"}\NormalTok{, }\StringTok{"All"}\NormalTok{), }
                    \StringTok{"Treatment"}\NormalTok{)}\SpecialCharTok{+}
  \FunctionTok{theme\_bw}\NormalTok{() }\SpecialCharTok{+} 
  \FunctionTok{ylab}\NormalTok{(}\StringTok{""}\NormalTok{) }\SpecialCharTok{+} 
  \FunctionTok{xlab}\NormalTok{(}\StringTok{"Pearson\textquotesingle{}s rho"}\NormalTok{)}
\end{Highlighting}
\end{Shaded}

\includegraphics{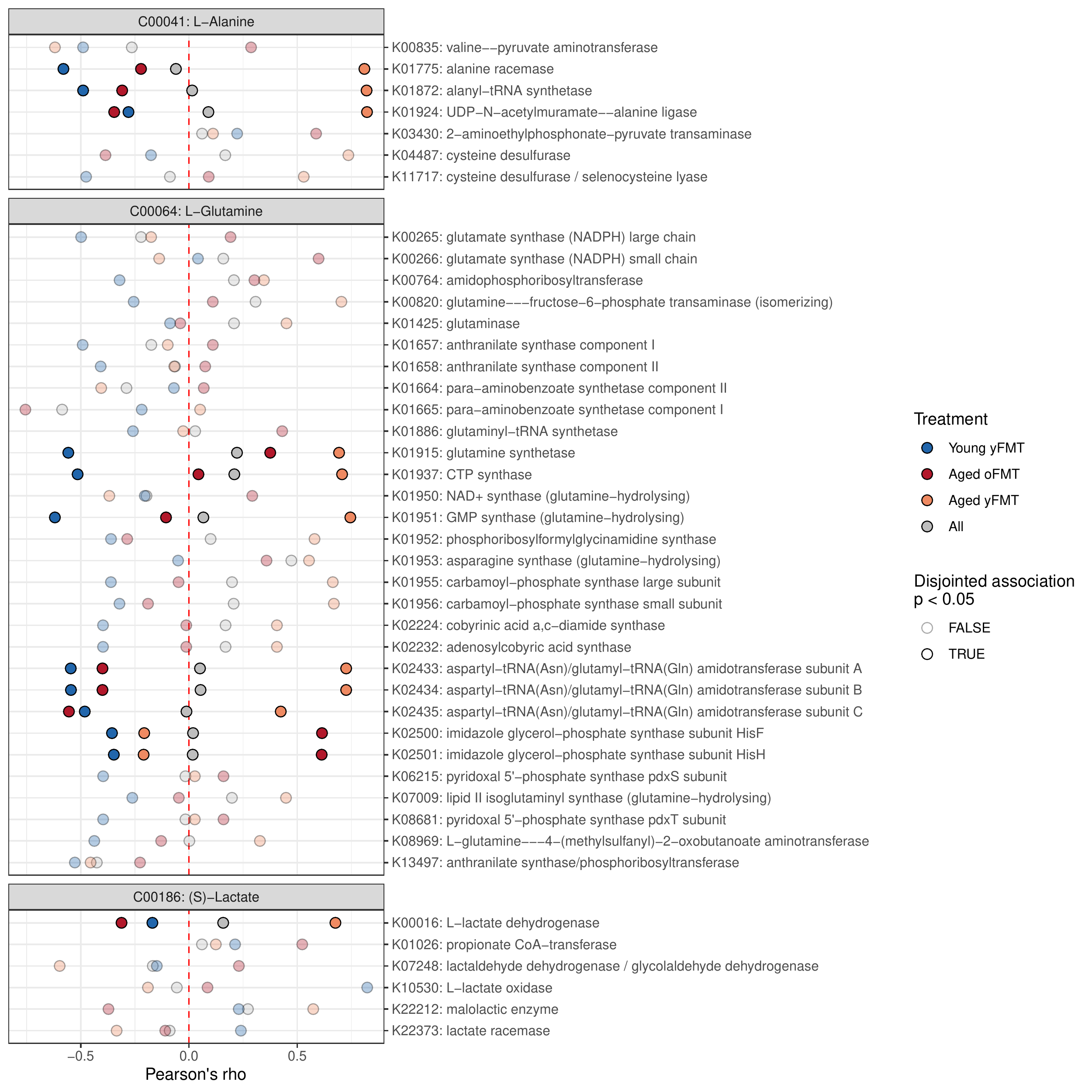}

Here, we can see the per-group correlations between metabolite-function
pairs in terms of Pearson's correlation coefficient on the x-axis.
Opaque points indicate significantly disjointed associations, meaning
that these associations have a significantly different slope between
groups.

\begin{center}\rule{0.5\linewidth}{0.5pt}\end{center}

\newpage

\hypertarget{computing-functional-modules}{%
\section{3. Computing functional
modules}\label{computing-functional-modules}}

Functional modules such as the Gut Brain modules shown here can be a
valuable framework to investigate your data through. One major benefit
of this framework is that you greatly reduce your search-space to
specific functional pathways you're interested in. This will allow you
to greatly save on statistical tests, which in turn will help save
statistical power when accounting for FDR.

Here, we will demonstrate how we got to the Gut Brain modules from our
the schizophrenia data set. Be warned that the first few operations here
are computationally very expensive and should be performed on a server
or similar if possible.

\hypertarget{gathering-and-preparing-our-data-1}{%
\subsection{Gathering and preparing our
data}\label{gathering-and-preparing-our-data-1}}

First, let's download the necessary data from
\texttt{curatedMicrobiomeData}

\hypertarget{code-chunk-download-and-pre-process-microbiome-data}{%
\subsubsection{Code chunk: Download and pre-process microbiome
data}\label{code-chunk-download-and-pre-process-microbiome-data}}

\begin{Shaded}
\begin{Highlighting}[]
\CommentTok{\#Load the relevant libraries}
\FunctionTok{library}\NormalTok{(curatedMetagenomicData)}
\FunctionTok{library}\NormalTok{(SummarizedExperiment)}

\CommentTok{\#Define what data we\textquotesingle{}re interested in. The "|" sign here signifies that we want both.}
\NormalTok{query }\OtherTok{\textless{}{-}} \StringTok{"2021{-}03{-}31.ZhuF\_2020.relative\_abundance|2021{-}03{-}31.ZhuF\_2020.gene\_families"}

\CommentTok{\#Download the specified data. This will take time. }
\NormalTok{ZhuF }\OtherTok{\textless{}{-}} \FunctionTok{curatedMetagenomicData}\NormalTok{(query, }\AttributeTok{counts =}\NormalTok{ T, }\AttributeTok{dryrun =}\NormalTok{ F)}

\CommentTok{\#Extract the relevant data from complex SummarizedExperiment objects}
\NormalTok{Zhu\_F\_gene\_families }\OtherTok{=} \FunctionTok{as.matrix}\NormalTok{( SummarizedExperiment}\SpecialCharTok{::}\FunctionTok{assay}\NormalTok{(ZhuF[[}\DecValTok{1}\NormalTok{]]))     }\CommentTok{\#Functions}
\NormalTok{Zhu\_F\_microbiome    }\OtherTok{=} \FunctionTok{data.frame}\NormalTok{(SummarizedExperiment}\SpecialCharTok{::}\FunctionTok{assay}\NormalTok{(ZhuF[[}\DecValTok{2}\NormalTok{]]))    }\CommentTok{\#Taxa}
\NormalTok{Zhu\_F\_metadata      }\OtherTok{=} \FunctionTok{data.frame}\NormalTok{(SummarizedExperiment}\SpecialCharTok{::}\FunctionTok{colData}\NormalTok{(ZhuF[[}\DecValTok{2}\NormalTok{]]))  }\CommentTok{\#Metadata}

\NormalTok{genus\_lv\_counts }\OtherTok{\textless{}{-}}\NormalTok{ Zhu\_F\_microbiome }\SpecialCharTok{\%\textgreater{}\%} 
  \FunctionTok{rownames\_to\_column}\NormalTok{(}\StringTok{"X"}\NormalTok{) }\SpecialCharTok{\%\textgreater{}\%}
  
  \CommentTok{\#Collapse to genus level}
  \FunctionTok{mutate}\NormalTok{(}\AttributeTok{X =} \FunctionTok{str\_remove}\NormalTok{(X, }\StringTok{"}\SpecialCharTok{\textbackslash{}\textbackslash{}}\StringTok{|s\_\_.*"}\NormalTok{)) }\SpecialCharTok{\%\textgreater{}\%} 
  \FunctionTok{group\_by}\NormalTok{(X) }\SpecialCharTok{\%\textgreater{}\%} 
  \FunctionTok{summarise}\NormalTok{(}\FunctionTok{across}\NormalTok{(}\FunctionTok{everything}\NormalTok{(), sum)) }\SpecialCharTok{\%\textgreater{}\%} 
  \FunctionTok{ungroup}\NormalTok{() }\SpecialCharTok{\%\textgreater{}\%} 
  
  \CommentTok{\#Clean up the names}
  \FunctionTok{mutate}\NormalTok{(}\AttributeTok{X =} \FunctionTok{str\_remove}\NormalTok{(X, }\StringTok{".*}\SpecialCharTok{\textbackslash{}\textbackslash{}}\StringTok{|f\_\_"}\NormalTok{)) }\SpecialCharTok{\%\textgreater{}\%} 
  \FunctionTok{mutate}\NormalTok{(}\AttributeTok{X =} \FunctionTok{str\_replace}\NormalTok{(X, }\StringTok{"}\SpecialCharTok{\textbackslash{}\textbackslash{}}\StringTok{|g\_\_"}\NormalTok{, }\StringTok{"\_"}\NormalTok{)) }\SpecialCharTok{\%\textgreater{}\%} 
  \FunctionTok{mutate}\NormalTok{(}\AttributeTok{X =} \FunctionTok{str\_replace}\NormalTok{(X, }\StringTok{"Clostridiales\_unclassified\_"}\NormalTok{, }\StringTok{"Clostridiales\_"}\NormalTok{)) }\SpecialCharTok{\%\textgreater{}\%} 
  \FunctionTok{filter}\NormalTok{(X }\SpecialCharTok{!=} \StringTok{"Clostridiales\_Clostridiales\_unclassified"}\NormalTok{) }\SpecialCharTok{\%\textgreater{}\%} 
  
  \CommentTok{\#Restore row.names}
  \FunctionTok{column\_to\_rownames}\NormalTok{(}\StringTok{"X"}\NormalTok{)}

\CommentTok{\#Should be identical to file used}
\NormalTok{waldo}\SpecialCharTok{::}\FunctionTok{compare}\NormalTok{(}\FunctionTok{sort}\NormalTok{(}\FunctionTok{row.names}\NormalTok{(genus\_lv\_counts)), }\FunctionTok{sort}\NormalTok{(}\FunctionTok{row.names}\NormalTok{(counts)))}

\CommentTok{\#Now that we have our data, we can write the individual tables to csv files. }
\CommentTok{\#We\textquotesingle{}ll focus on the gene families here, they\textquotesingle{}re necessary to compute Gut Brain modules.}

\FunctionTok{write.csv}\NormalTok{(Zhu\_F\_gene\_families, }\AttributeTok{file =} \StringTok{"uniref90.csv"}\NormalTok{)}
\FunctionTok{write.csv}\NormalTok{(Zhu\_F\_microbiome,    }\AttributeTok{file =} \StringTok{"counts.csv"}\NormalTok{)}
\end{Highlighting}
\end{Shaded}

\hypertarget{convert-to-kegg-orthologues}{%
\subsection{Convert to KEGG
orthologues}\label{convert-to-kegg-orthologues}}

\texttt{Zhu\_F\_gene\_families} contains the functional microbiome in
terms of uniref90. For functional module analysis, we typically want to
get to KEGG orthologues (KOs). Because \texttt{curatedMetagenomicData}
gives essentially biobakery output, we can use commands from the
python-based \texttt{HUMAnN3} pipeline to translate our uniref90 table
to KEGG orthologues. We will not go deeply into this, but see the
excellent documentation here:
\url{https://github.com/biobakery/humann\#humann_regroup_table}

\textbf{The next snippet is not R code, but rather Bash code.}

\hypertarget{code-chunk-convert-functional-table-to-kegg-orthologues}{%
\subsubsection{Code chunk: Convert functional table to KEGG
orthologues}\label{code-chunk-convert-functional-table-to-kegg-orthologues}}

\begin{Shaded}
\begin{Highlighting}[]

\CommentTok{\#First, we may want to change our uniref90 file so that it uses tabs instead of commas}
\FunctionTok{sed} \AttributeTok{{-}E} \StringTok{\textquotesingle{}s/("([\^{}"]*)")?,/\textbackslash{}2\textbackslash{}t/g\textquotesingle{}}\NormalTok{ uniref90.csv  }\OperatorTok{\textgreater{}}\NormalTok{ uniref90.tsv}

\CommentTok{\#Then, let\textquotesingle{}s use humann\_regroup\_table from HUMAnN3 to convert to KEGG orthologues:}
\ExtensionTok{humann\_regroup\_table} \AttributeTok{{-}i}\NormalTok{ uniref90.tsv }\AttributeTok{{-}g}\NormalTok{ uniref90\_ko }\AttributeTok{{-}o}\NormalTok{ guidebook\_out\_KOs.tsv}
\end{Highlighting}
\end{Shaded}

\newpage

\hypertarget{compute-functional-modules}{%
\subsection{Compute functional
modules}\label{compute-functional-modules}}

Now that we have our functional microbiome in terms of KEGG orthologues,
we can load them back into R. An added benefit is that the KEGG
orthologues table is much smaller than the uniref90 table, so we can
deal with it on our computer locally if we so choose. In order to do
this, we will require the \texttt{omixer-rpmR} library which can be
found of github. If you're working with functional inference data from a
16S experiment, such as output from \texttt{PICRUSt2}, you should be
able to read the table in and compute functional modules starting at
this step. The file you're looking for would be called something like
\texttt{pred\_metagenome\_unstrat.tsv} in that case.

\hypertarget{code-chunk-generate-functional-modules}{%
\subsubsection{Code chunk: Generate functional
modules}\label{code-chunk-generate-functional-modules}}

\begin{Shaded}
\begin{Highlighting}[]
\CommentTok{\#Load the required package}
\FunctionTok{library}\NormalTok{(omixerRpm) }\CommentTok{\#devtools::install\_github("omixer/omixer{-}rpmR")}

\CommentTok{\#Load the KEGG orthologue table into R. }
\CommentTok{\#Note that KEGG orthologue names should be the in first column, not the row names.  }
\NormalTok{KOs }\OtherTok{\textless{}{-}} \FunctionTok{read.delim}\NormalTok{(}\StringTok{"guidebook\_out\_KOs.tsv"}\NormalTok{, }\AttributeTok{header =}\NormalTok{ T)}

\CommentTok{\#listDB will tell you which databases are available to annotate the functional data with. }
\FunctionTok{listDB}\NormalTok{()}

\CommentTok{\#Pick the most recent GBM database}
\NormalTok{db }\OtherTok{\textless{}{-}} \FunctionTok{loadDB}\NormalTok{(}\AttributeTok{name =} \StringTok{"GBMs.v1.0"}\NormalTok{)}

\CommentTok{\#Calculate GBM abundance and convert the output to a nice data.frame}
\NormalTok{GBMs }\OtherTok{\textless{}{-}} \FunctionTok{rpm}\NormalTok{(}\AttributeTok{x =}\NormalTok{ KOs,  }\AttributeTok{module.db =}\NormalTok{ db)}
\NormalTok{GBMs }\OtherTok{\textless{}{-}} \FunctionTok{asDataFrame}\NormalTok{(GBMs, }\StringTok{"abundance"}\NormalTok{)}

\CommentTok{\#Write the file to a csv to save it. }
\FunctionTok{write.csv}\NormalTok{(GBMs, }\AttributeTok{file =} \StringTok{"GBMs\_guidebook.csv"}\NormalTok{)}

\CommentTok{\#While we\textquotesingle{}re at it, let\textquotesingle{}s do GMMs too: First check the names of the available databases:}
\FunctionTok{listDB}\NormalTok{()}

\CommentTok{\#Pick the most recent GMM database}
\NormalTok{db }\OtherTok{\textless{}{-}} \FunctionTok{loadDB}\NormalTok{(}\AttributeTok{name =} \StringTok{"GMMs.v1.07"}\NormalTok{)}

\CommentTok{\#Calculate GMM abundance and convert the output to a nice data.frame}
\NormalTok{GMMs }\OtherTok{\textless{}{-}} \FunctionTok{rpm}\NormalTok{(}\AttributeTok{x =}\NormalTok{ KOs,  }\AttributeTok{module.db =}\NormalTok{ db)}
\NormalTok{GMMs }\OtherTok{\textless{}{-}} \FunctionTok{asDataFrame}\NormalTok{(GMMs, }\StringTok{"abundance"}\NormalTok{)}

\CommentTok{\#Write the file to a csv to save it. }
\FunctionTok{write.csv}\NormalTok{(GMMs, }\AttributeTok{file =} \StringTok{"GMMs\_guidebook.csv"}\NormalTok{)}
\end{Highlighting}
\end{Shaded}

And the resulting files are ready for statistical analysis! I would like
to note here that it's also possible to perform a stratified functional
module analysis, where the contribution of each taxon to each functional
module is also considered. However, this explosively increases the
dimensionality of your data (i.e.~you get way more rows in our case). I
would only recommend doing this as a targeted analysis as the power of
any statistical tests will suffer greatly and the results will be almost
impossible to interpret.

\begin{center}\rule{0.5\linewidth}{0.5pt}\end{center}

\newpage

\hypertarget{volatility-analysis}{%
\section{4. Volatility Analysis}\label{volatility-analysis}}

Volatility refers to the degree of instability (or change over time) in
the microbiome. High volatility, i.e.~an unstable microbiome, has been
associated with an exaggerated stress response and conditions like IBS.
Here, we'll demonstrate how one would go about calculating volatility in
a real data set. Volatility analysis requires at least two time points
per sample. Because of this, we cannot use the schizophrenia data set
which only features single snapshots of the microbiome.

We'll be taking a look at the datasets used in the original Volatility
paper: \emph{Volatility as a Concept to Understand the Impact of Stress
on the Microbiome} (DOI: 10.1016/j.psyneuen.2020.105047). Very briefly,
mice were separated into two groups: Control and Stress. Faecal samples
were taken twice, with a 10-day period in between. In this 10-day
period, the mice in the Stress group were subjected to daily social
defeat stress, whereas the Control mice were left alone. When we
compared the degree of change in the microbiome (i.e.~Volatility)
between the two groups of mice, the Stressed mice consistently displayed
higher levels of volatility than the control mice. Our reviewers asked
us to replicate the experiment and we did so. The two cohorts are
labeled discovery and validation. This data has been included in the
\texttt{volatility} library on github for easy access purposes.

Traditionally, microbiome studies featuring high-throughput sequencing
data only consider a single time point. However, there is utility in
considering microbiomes as dynamic microbial ecosystems that change over
time. By measuring the microbiome longitudinally and computing
volatility, additional information can be revealed that would otherwise
be missed. For instance, in the original volatility study, we found that
volatility after stress is positively associated with severity of the
stress response, including in terms of behaviour and
hypothalamic-pituitary-adrenal (HPA) axis activity, both in mice and in
humans.

\hypertarget{setup}{%
\subsection{Setup}\label{setup}}

OK, now let's get started.

\hypertarget{code-chunk-load-data}{%
\subsubsection{Code chunk: Load data}\label{code-chunk-load-data}}

\begin{Shaded}
\begin{Highlighting}[]
\CommentTok{\#Install and load volatility library}
\FunctionTok{library}\NormalTok{(volatility)     }\CommentTok{\#devtools::install\_github("thomazbastiaanssen/volatility")}

\CommentTok{\#Load tidyverse to wrangle and plot results.}
\FunctionTok{library}\NormalTok{(tidyverse)}

\CommentTok{\#Load example data + metadata from the volatility study.}
\FunctionTok{data}\NormalTok{(volatility\_data)}
\end{Highlighting}
\end{Shaded}

\newpage

\hypertarget{considering-our-input-data}{%
\subsection{Considering our input
data}\label{considering-our-input-data}}

The main \texttt{volatility} function does all the heavy lifting here.
It expects two arguments: The first argument is \texttt{counts}, a
microbiome feature count table, with columns as samples and rows and
features. The \texttt{vola\_genus\_table} object is an example of an
appropriately formatted count table.

\hypertarget{code-chunk-examine-required-data-format}{%
\subsubsection{Code chunk: Examine required data
format}\label{code-chunk-examine-required-data-format}}

\begin{Shaded}
\begin{Highlighting}[]
\NormalTok{vola\_genus\_table[}\DecValTok{4}\SpecialCharTok{:}\DecValTok{9}\NormalTok{,}\DecValTok{1}\SpecialCharTok{:}\DecValTok{2}\NormalTok{]}
\end{Highlighting}
\end{Shaded}

\begin{verbatim}
##                               Validation_Pre_Control_1 Validation_Pre_Control_2
## Atopobiaceae_Olsenella                               0                        0
## Coriobacteriaceae_Collinsella                        0                        0
## Eggerthellaceae_DNF00809                           102                       47
## Eggerthellaceae_Enterorhabdus                       53                      114
## Eggerthellaceae_Parvibacter                         21                       20
## Bacteroidaceae_Bacteroides                         616                      453
\end{verbatim}

The second arguent is \texttt{metadata}, a vector in the same order as
the count table, denoting which samples are from the same source. The
column \texttt{ID} in \texttt{vola\_metadata} is appropriate for this.

\hypertarget{code-chunk-examine-required-metadata-format}{%
\subsubsection{Code chunk: Examine required metadata
format}\label{code-chunk-examine-required-metadata-format}}

\begin{Shaded}
\begin{Highlighting}[]
\FunctionTok{head}\NormalTok{(vola\_metadata, }\DecValTok{5}\NormalTok{)}
\end{Highlighting}
\end{Shaded}

\begin{verbatim}
##                  sample_ID     cohort timepoint treatment ID
## 1 Validation_Pre_Control_1 Validation       Pre   Control  1
## 2 Validation_Pre_Control_2 Validation       Pre   Control  2
## 3 Validation_Pre_Control_3 Validation       Pre   Control  3
## 4 Validation_Pre_Control_4 Validation       Pre   Control  4
## 5 Validation_Pre_Control_5 Validation       Pre   Control  5
\end{verbatim}

\hypertarget{code-chunk-prepare-the-data-for-plotting}{%
\subsubsection{Code chunk: Prepare the data for
plotting}\label{code-chunk-prepare-the-data-for-plotting}}

\begin{Shaded}
\begin{Highlighting}[]
\CommentTok{\#This part should feel very reminiscent of what we did in chapters 1 \& 3. }
\NormalTok{counts  }\OtherTok{\textless{}{-}}\NormalTok{ vola\_genus\_table[,vola\_metadata}\SpecialCharTok{$}\NormalTok{sample\_ID]}

\CommentTok{\#Fork off your count data so that you always have an untouched version handy.}
\NormalTok{genus   }\OtherTok{\textless{}{-}}\NormalTok{ counts}

\CommentTok{\#make sure our count data is all numbers}
\NormalTok{genus   }\OtherTok{\textless{}{-}} \FunctionTok{apply}\NormalTok{(genus,}\FunctionTok{c}\NormalTok{(}\DecValTok{1}\NormalTok{,}\DecValTok{2}\NormalTok{),}\ControlFlowTok{function}\NormalTok{(x) }\FunctionTok{as.numeric}\NormalTok{(}\FunctionTok{as.character}\NormalTok{(x)))}

\CommentTok{\#Remove features with prevalence \textless{} 10\% in two steps:}
\CommentTok{\#First, determine how often every feature is absent in a sample}
\NormalTok{n\_zeroes }\OtherTok{\textless{}{-}} \FunctionTok{rowSums}\NormalTok{(genus }\SpecialCharTok{==} \DecValTok{0}\NormalTok{)}

\CommentTok{\#Then, remove features that are absent in more than your threshold (90\% in this case).}
\NormalTok{genus    }\OtherTok{\textless{}{-}}\NormalTok{ genus[n\_zeroes }\SpecialCharTok{\textless{}=} \FunctionTok{round}\NormalTok{(}\FunctionTok{ncol}\NormalTok{(genus) }\SpecialCharTok{*} \FloatTok{0.90}\NormalTok{),]}

\CommentTok{\#Perform a CLR transformation}
\NormalTok{genus.exp }\OtherTok{\textless{}{-}} \FunctionTok{clr\_c}\NormalTok{(genus)}

\CommentTok{\#Apply the base R principal component analysis function on our CLR{-}transformed data.}
\NormalTok{data.a.pca  }\OtherTok{\textless{}{-}} \FunctionTok{prcomp}\NormalTok{(}\FunctionTok{t}\NormalTok{(genus.exp))}
\end{Highlighting}
\end{Shaded}

\newpage

\hypertarget{plot-data}{%
\subsection{Plot data}\label{plot-data}}

\hypertarget{code-chunk-plot-longitudinal-pca}{%
\subsubsection{Code chunk: Plot longitudinal
PCA}\label{code-chunk-plot-longitudinal-pca}}

\begin{Shaded}
\begin{Highlighting}[]
\CommentTok{\#Extract the amount of variance the first four components explain for plotting. }
\NormalTok{pc1 }\OtherTok{\textless{}{-}} \FunctionTok{round}\NormalTok{(data.a.pca}\SpecialCharTok{$}\NormalTok{sdev[}\DecValTok{1}\NormalTok{]}\SpecialCharTok{\^{}}\DecValTok{2}\SpecialCharTok{/}\FunctionTok{sum}\NormalTok{(data.a.pca}\SpecialCharTok{$}\NormalTok{sdev}\SpecialCharTok{\^{}}\DecValTok{2}\NormalTok{),}\DecValTok{4}\NormalTok{) }\SpecialCharTok{*} \DecValTok{100}
\NormalTok{pc2 }\OtherTok{\textless{}{-}} \FunctionTok{round}\NormalTok{(data.a.pca}\SpecialCharTok{$}\NormalTok{sdev[}\DecValTok{2}\NormalTok{]}\SpecialCharTok{\^{}}\DecValTok{2}\SpecialCharTok{/}\FunctionTok{sum}\NormalTok{(data.a.pca}\SpecialCharTok{$}\NormalTok{sdev}\SpecialCharTok{\^{}}\DecValTok{2}\NormalTok{),}\DecValTok{4}\NormalTok{) }\SpecialCharTok{*} \DecValTok{100}
\NormalTok{pc3 }\OtherTok{\textless{}{-}} \FunctionTok{round}\NormalTok{(data.a.pca}\SpecialCharTok{$}\NormalTok{sdev[}\DecValTok{3}\NormalTok{]}\SpecialCharTok{\^{}}\DecValTok{2}\SpecialCharTok{/}\FunctionTok{sum}\NormalTok{(data.a.pca}\SpecialCharTok{$}\NormalTok{sdev}\SpecialCharTok{\^{}}\DecValTok{2}\NormalTok{),}\DecValTok{4}\NormalTok{) }\SpecialCharTok{*} \DecValTok{100}
\NormalTok{pc4 }\OtherTok{\textless{}{-}} \FunctionTok{round}\NormalTok{(data.a.pca}\SpecialCharTok{$}\NormalTok{sdev[}\DecValTok{4}\NormalTok{]}\SpecialCharTok{\^{}}\DecValTok{2}\SpecialCharTok{/}\FunctionTok{sum}\NormalTok{(data.a.pca}\SpecialCharTok{$}\NormalTok{sdev}\SpecialCharTok{\^{}}\DecValTok{2}\NormalTok{),}\DecValTok{4}\NormalTok{) }\SpecialCharTok{*} \DecValTok{100}

\CommentTok{\#Extract the scores for every sample for the first four components for plotting. }
\NormalTok{pca  }\OtherTok{=} \FunctionTok{data.frame}\NormalTok{(}\AttributeTok{PC1 =}\NormalTok{ data.a.pca}\SpecialCharTok{$}\NormalTok{x[,}\DecValTok{1}\NormalTok{], }
                  \AttributeTok{PC2 =}\NormalTok{ data.a.pca}\SpecialCharTok{$}\NormalTok{x[,}\DecValTok{2}\NormalTok{], }
                  \AttributeTok{PC3 =}\NormalTok{ data.a.pca}\SpecialCharTok{$}\NormalTok{x[,}\DecValTok{3}\NormalTok{], }
                  \AttributeTok{PC4 =}\NormalTok{ data.a.pca}\SpecialCharTok{$}\NormalTok{x[,}\DecValTok{4}\NormalTok{])}

\CommentTok{\#Add relevant information from the metadata. }
\CommentTok{\#Note that ID here refers to the mouse ID, not the sample ID. }
\NormalTok{pca}\SpecialCharTok{$}\NormalTok{ID                  }\OtherTok{=}\NormalTok{ vola\_metadata}\SpecialCharTok{$}\NormalTok{ID}
\NormalTok{pca}\SpecialCharTok{$}\NormalTok{Legend              }\OtherTok{=}\NormalTok{ vola\_metadata}\SpecialCharTok{$}\NormalTok{treatment}
\NormalTok{pca}\SpecialCharTok{$}\NormalTok{Timepoint           }\OtherTok{=}\NormalTok{ vola\_metadata}\SpecialCharTok{$}\NormalTok{timepoint}
\NormalTok{pca}\SpecialCharTok{$}\NormalTok{Cohort              }\OtherTok{=}\NormalTok{ vola\_metadata}\SpecialCharTok{$}\NormalTok{cohort}

\CommentTok{\#Plot the first two components of the PCA.}
\FunctionTok{ggplot}\NormalTok{(pca, }\FunctionTok{aes}\NormalTok{(}\AttributeTok{x       =}\NormalTok{ PC1, }
                \AttributeTok{y       =}\NormalTok{ PC2, }
                \AttributeTok{fill    =}\NormalTok{ Legend,}
                \AttributeTok{colour  =}\NormalTok{ Legend,}
                \AttributeTok{shape   =}\NormalTok{ Timepoint, }
                \AttributeTok{group   =}\NormalTok{ ID)) }\SpecialCharTok{+}  
  \CommentTok{\#Add a line first, it will link points that share an ID. }
  \FunctionTok{geom\_line}\NormalTok{() }\SpecialCharTok{+}
  
  \CommentTok{\#Then add the points.}
  \FunctionTok{geom\_point}\NormalTok{(}\AttributeTok{size =} \DecValTok{3}\NormalTok{, }\AttributeTok{col =} \StringTok{"black"}\NormalTok{) }\SpecialCharTok{+} 
  
  \CommentTok{\#Plot the two cohorts separately. }
  \FunctionTok{facet\_wrap}\NormalTok{(}\SpecialCharTok{\textasciitilde{}}\NormalTok{Cohort, }\AttributeTok{scales =} \StringTok{"free\_x"}\NormalTok{, }\AttributeTok{strip.position =} \StringTok{"top"}\NormalTok{) }\SpecialCharTok{+}
  
  \CommentTok{\#Improve appearance.}
  \FunctionTok{scale\_fill\_manual}\NormalTok{(  }\AttributeTok{values =} \FunctionTok{c}\NormalTok{(}\StringTok{"Control"}  \OtherTok{=} \StringTok{"\#1f78b4"}\NormalTok{, }
                                 \StringTok{"Stress"}   \OtherTok{=} \StringTok{"\#e31a1c"}\NormalTok{)) }\SpecialCharTok{+} 
  \FunctionTok{scale\_colour\_manual}\NormalTok{(}\AttributeTok{values =} \FunctionTok{c}\NormalTok{(}\StringTok{"Control"}  \OtherTok{=} \StringTok{"\#3690c0"}\NormalTok{, }
                                 \StringTok{"Stress"}   \OtherTok{=} \StringTok{"\#cb181d"}\NormalTok{)) }\SpecialCharTok{+} 
  \FunctionTok{scale\_shape\_manual}\NormalTok{( }\AttributeTok{values =} \FunctionTok{c}\NormalTok{(}\StringTok{"Pre"}  \OtherTok{=} \DecValTok{21}\NormalTok{, }
                                \StringTok{"Post"} \OtherTok{=} \DecValTok{24}\NormalTok{)) }\SpecialCharTok{+}  
  \FunctionTok{theme\_bw}\NormalTok{() }\SpecialCharTok{+}
  \FunctionTok{xlab}\NormalTok{(}\FunctionTok{paste}\NormalTok{(}\StringTok{"PC1: "}\NormalTok{, pc1,  }\StringTok{"\%"}\NormalTok{, }\AttributeTok{sep=}\StringTok{""}\NormalTok{)) }\SpecialCharTok{+} 
  \FunctionTok{ylab}\NormalTok{(}\FunctionTok{paste}\NormalTok{(}\StringTok{"PC2: "}\NormalTok{, pc2,  }\StringTok{"\%"}\NormalTok{, }\AttributeTok{sep=}\StringTok{""}\NormalTok{)) }\SpecialCharTok{+} 
  \FunctionTok{theme}\NormalTok{(}\AttributeTok{text =} \FunctionTok{element\_text}\NormalTok{(}\AttributeTok{size =} \DecValTok{12}\NormalTok{)) }\SpecialCharTok{+}
  \FunctionTok{guides}\NormalTok{(}\AttributeTok{fill =} \FunctionTok{guide\_legend}\NormalTok{(}\AttributeTok{override.aes =} \FunctionTok{list}\NormalTok{(}\AttributeTok{shape =} \DecValTok{22}\NormalTok{)))}
\end{Highlighting}
\end{Shaded}

\includegraphics{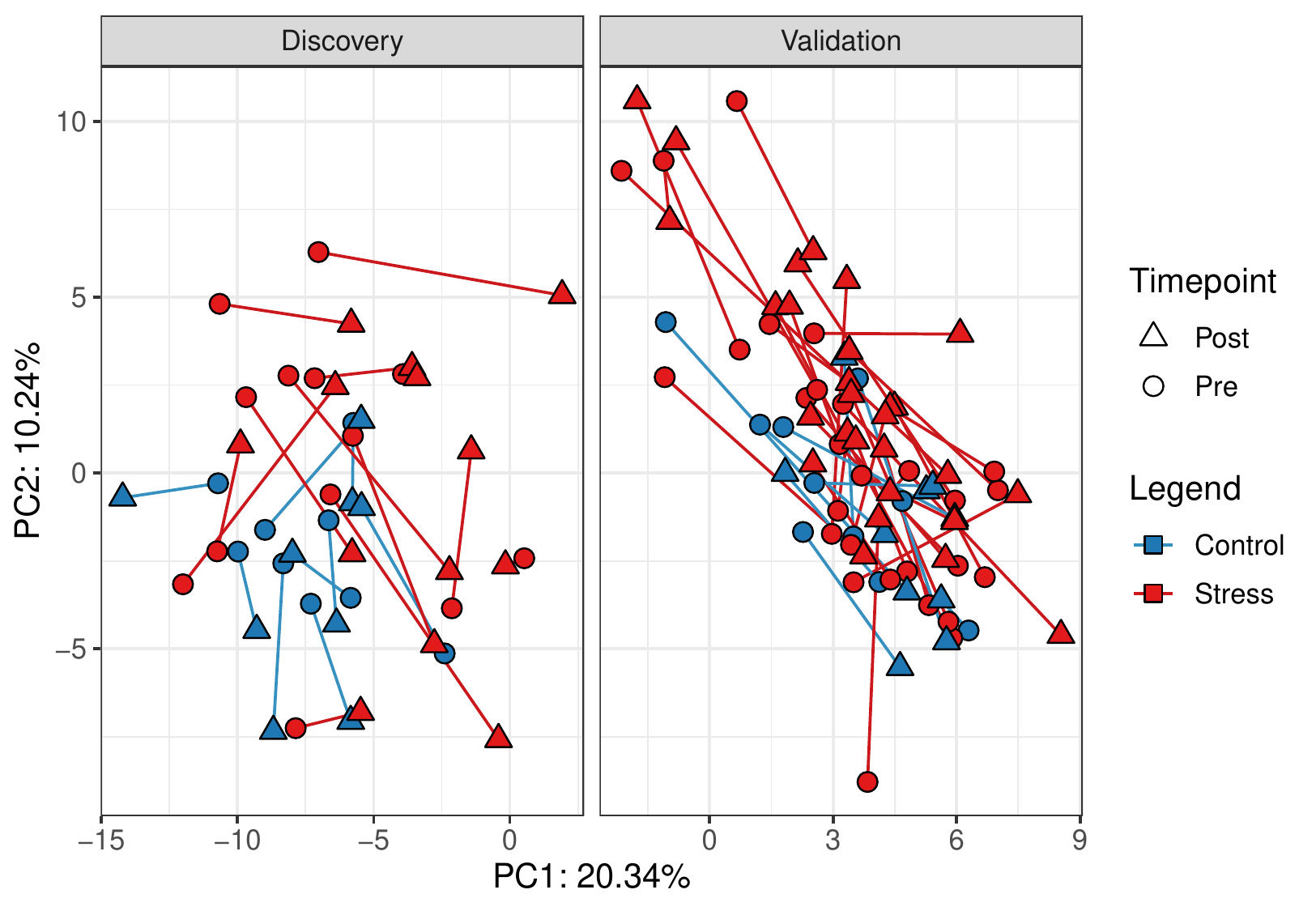}

We can see that points from the same mouse are connected by a line. It
looks like some lines are longer than others, implying that some
microbiomes have changed more than others over the 10 days. However,
We're only looking at about 30\% of the variance here, so it's hard to
say anything conclusive.

\hypertarget{compute-volatility}{%
\subsection{Compute volatility}\label{compute-volatility}}

Volatility between two samples can be easily calculated using the
titular \texttt{volatilty} function in the library by the same name.
Under the hood, volatility can be calculated as the euclidean distance
over CLR-transformed count data.

\hypertarget{code-chunk-calculate-volatility}{%
\subsubsection{Code chunk: Calculate
volatility}\label{code-chunk-calculate-volatility}}

\begin{Shaded}
\begin{Highlighting}[]
\NormalTok{vola\_out }\OtherTok{\textless{}{-}} \FunctionTok{volatility}\NormalTok{(}\AttributeTok{counts =}\NormalTok{ genus, }\AttributeTok{metadata =}\NormalTok{ vola\_metadata}\SpecialCharTok{$}\NormalTok{ID)}

\FunctionTok{head}\NormalTok{(vola\_out)}
\end{Highlighting}
\end{Shaded}

\begin{verbatim}
##   ID volatility
## 1  1  13.275363
## 2 10  12.979717
## 3 11  15.987468
## 4 13  14.902772
## 5 14   8.094823
## 6 15   9.506299
\end{verbatim}

The output of the main \texttt{volatility} function is a data.frame with
two columns. \texttt{ID} corresponds to the pairs of samples passed on
in the \texttt{metadata} argument, whereas \texttt{volatility} shows the
measured volatility between those samples in terms of Aitchison distance
(Euclidean distance of CLR-transformed counts).

\newpage

\hypertarget{plot-the-results-1}{%
\subsection{Plot the results}\label{plot-the-results-1}}

\hypertarget{code-chunk-plot-volatility}{%
\subsubsection{Code chunk: Plot
volatility}\label{code-chunk-plot-volatility}}

\begin{Shaded}
\begin{Highlighting}[]
\NormalTok{vola\_out }\SpecialCharTok{\%\textgreater{}\%}
  \CommentTok{\#Merge the volatilty output with the rest of the metadata using the shared "ID" column.}
  \FunctionTok{left\_join}\NormalTok{(vola\_metadata[vola\_metadata}\SpecialCharTok{$}\NormalTok{timepoint }\SpecialCharTok{==} \StringTok{"Pre"}\NormalTok{,], }\StringTok{"ID"}\NormalTok{) }\SpecialCharTok{\%\textgreater{}\%}
  
  \CommentTok{\#Pipe into ggplot.}
  \FunctionTok{ggplot}\NormalTok{(}\FunctionTok{aes}\NormalTok{(}\AttributeTok{x =}\NormalTok{ treatment, }\AttributeTok{y =}\NormalTok{ volatility, }\AttributeTok{fill =}\NormalTok{ treatment)) }\SpecialCharTok{+}
  
  \CommentTok{\#Define geoms, boxplots overlayed with data points in this case.}
  \FunctionTok{geom\_boxplot}\NormalTok{(}\AttributeTok{alpha =} \DecValTok{1}\SpecialCharTok{/}\DecValTok{2}\NormalTok{, }\AttributeTok{coef =} \DecValTok{10}\NormalTok{)}\SpecialCharTok{+}
  \FunctionTok{geom\_point}\NormalTok{(}\AttributeTok{shape =} \DecValTok{21}\NormalTok{) }\SpecialCharTok{+}
  
  \CommentTok{\#Split the plot by cohort.}
  \FunctionTok{facet\_wrap}\NormalTok{(}\SpecialCharTok{\textasciitilde{}}\NormalTok{cohort) }\SpecialCharTok{+}
  
  \CommentTok{\#Tweak appearance.}
  \FunctionTok{scale\_fill\_manual}\NormalTok{(}\AttributeTok{values =} \FunctionTok{c}\NormalTok{(}\StringTok{"Control"} \OtherTok{=} \StringTok{"\#3690c0"}\NormalTok{, }\StringTok{"Stress"}  \OtherTok{=} \StringTok{"\#cb181d"}\NormalTok{)) }\SpecialCharTok{+}
  \FunctionTok{theme\_bw}\NormalTok{() }\SpecialCharTok{+}
  \FunctionTok{xlab}\NormalTok{(}\StringTok{""}\NormalTok{) }\SpecialCharTok{+}
  \FunctionTok{ylab}\NormalTok{(}\StringTok{"Volatility (Aitchison distance)"}\NormalTok{)}
\end{Highlighting}
\end{Shaded}

\includegraphics{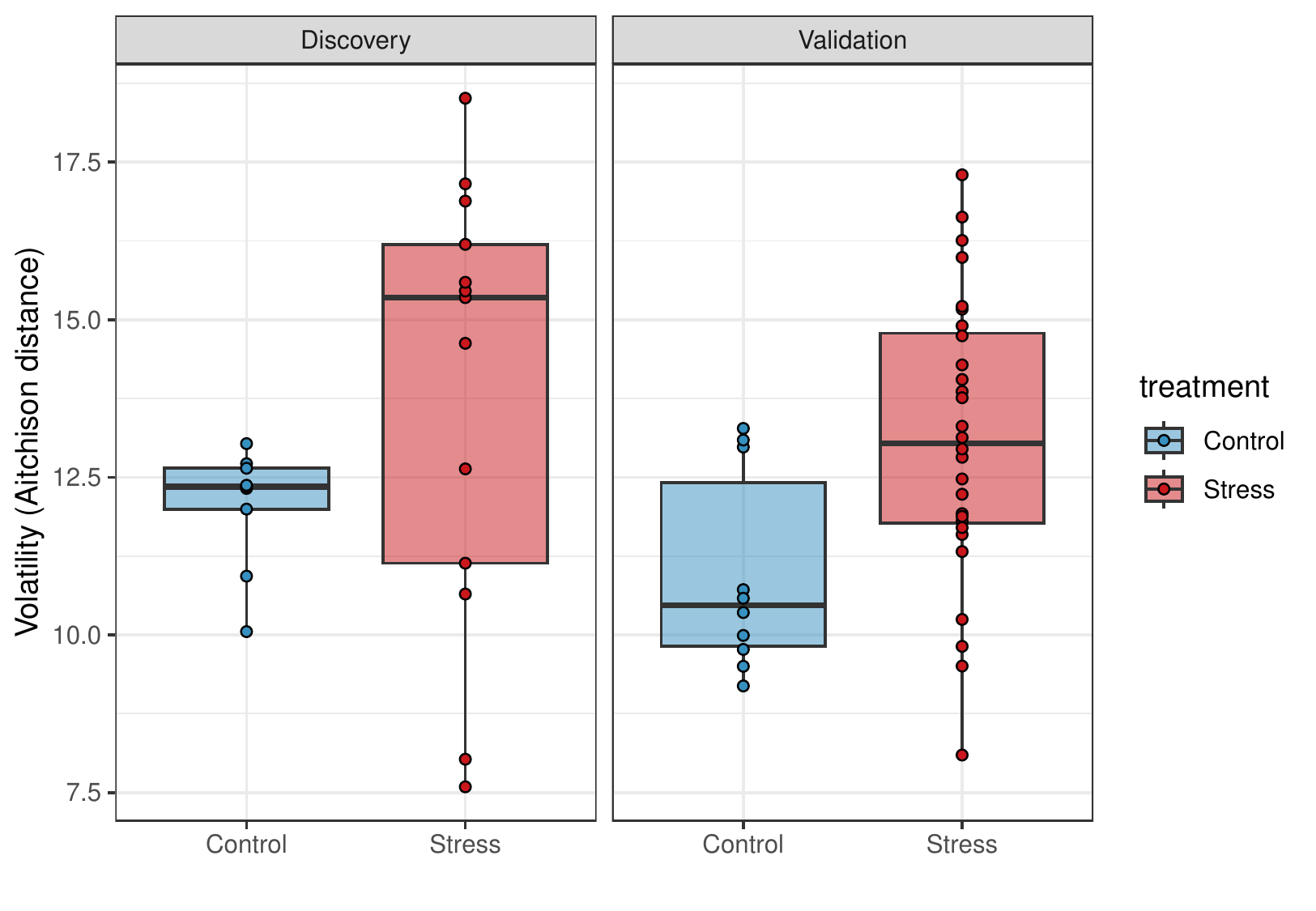}

\begin{center}\rule{0.5\linewidth}{0.5pt}\end{center}

\newpage

\hypertarget{discussion}{%
\section{5. Discussion}\label{discussion}}

Here, we have presented four separate primers for techniques from four
distinct topics, namely causal inference, multi-omics integration,
mesoscale analysis and temporal analysis, and how one may go about
applying them to microbiome-gut-brain axis experiments. While these
techniques and corresponding fields may seem unrelated, we are of the
opinion that all four will be essential to move the microbiome-gut-brain
axis field forward towards an ecology oriented, mechanistic
understanding.

As said in the discussion of part 1 of this perspective, this document
is just a template. Depending on the experimental setup, findings and
experimental questions, you may want to choose a differing approach.
Given the highly complex nature of microbiome data, one should ideally
avoid blindly applying models and pipelines without understanding what
they are doing. D.R. Cox is famously ascribed the statement:
\emph{``Most real life statistical problems have one or more nonstandard
features. There are no routine statistical questions; only questionable
statistical routines.''} We find this holds true for the microbiome as
well, even more so for these more advanced techniques than for the ones
presented in part 1 of this perspective piece. In particular, it is
crucial that we apply our biological knowledge to determine how exactly
to use these techniques. For instance, when performing a mediation
analysis in section 1, it is crucial to have a biological reason why one
would expect the effect of a variable on an outcome to be moderated by a
second variable. Similarly, with the multi-omics integration using
anansi in section 2, we're relying on pre-existing biological knowledge
in the form of the KEGG database to only investigate interactions
between metabolites and enzymatic functions that could take place,
rather than naively assessing all of them.

Clear communication, both in terms of describing and explaining our
methods as well as in terms of figure presentation, are essential for
the health of the field. Indeed, failing to do so can lead to confusion
among our peers. We hope that both aspiring and veteran
bioinformaticians will find our guide helpful. We have tried to model
this piece after what we would have loved to have access to ourselves
when we first set out to study the microbiome.

\begin{center}\rule{0.5\linewidth}{0.5pt}\end{center}

\newpage

\hypertarget{session-info}{%
\section{Session Info}\label{session-info}}

\begin{Shaded}
\begin{Highlighting}[]
\NormalTok{sessioninfo}\SpecialCharTok{::}\FunctionTok{session\_info}\NormalTok{()}
\end{Highlighting}
\end{Shaded}

\begin{verbatim}
## - Session info ---------------------------------------------------------------
##  setting  value
##  version  R version 4.2.2 Patched (2022-11-10 r83330)
##  os       Ubuntu 18.04.6 LTS
##  system   x86_64, linux-gnu
##  ui       X11
##  language en_IE:en
##  collate  en_IE.UTF-8
##  ctype    en_IE.UTF-8
##  tz       Europe/Dublin
##  date     2023-07-25
##  pandoc   2.19.2 @ /usr/lib/rstudio/resources/app/bin/quarto/bin/tools/ (via rmarkdown)
## 
## - Packages -------------------------------------------------------------------
##  package       * version    date (UTC) lib source
##  anansi        * 0.5.0      2023-04-25 [1] Github (thomazbastiaanssen/anansi@e188997)
##  assertthat      0.2.1      2019-03-21 [1] CRAN (R 4.2.0)
##  backports       1.4.1      2021-12-13 [1] CRAN (R 4.2.0)
##  base64enc       0.1-3      2015-07-28 [1] CRAN (R 4.2.0)
##  boot            1.3-28     2021-05-03 [4] CRAN (R 4.0.5)
##  broom           1.0.2      2022-12-15 [1] CRAN (R 4.2.1)
##  cellranger      1.1.0      2016-07-27 [1] CRAN (R 4.2.0)
##  checkmate       2.1.0      2022-04-21 [1] CRAN (R 4.2.0)
##  cli             3.6.0      2023-01-09 [1] CRAN (R 4.2.1)
##  cluster         2.1.4      2022-08-22 [4] CRAN (R 4.2.1)
##  codetools       0.2-19     2023-02-01 [4] CRAN (R 4.2.2)
##  colorspace      2.0-3      2022-02-21 [1] CRAN (R 4.2.0)
##  crayon          1.5.2      2022-09-29 [1] CRAN (R 4.2.1)
##  data.table      1.14.6     2022-11-16 [1] CRAN (R 4.2.1)
##  DBI             1.1.3      2022-06-18 [1] CRAN (R 4.2.0)
##  dbplyr          2.3.0      2023-01-16 [1] CRAN (R 4.2.1)
##  deldir          1.0-6      2021-10-23 [1] CRAN (R 4.2.1)
##  digest          0.6.31     2022-12-11 [1] CRAN (R 4.2.1)
##  dplyr         * 1.0.10     2022-09-01 [1] CRAN (R 4.2.1)
##  ellipsis        0.3.2      2021-04-29 [1] CRAN (R 4.2.0)
##  evaluate        0.20       2023-01-17 [1] CRAN (R 4.2.1)
##  fansi           1.0.3      2022-03-24 [1] CRAN (R 4.2.0)
##  farver          2.1.1      2022-07-06 [1] CRAN (R 4.2.1)
##  fastmap         1.1.0      2021-01-25 [1] CRAN (R 4.2.0)
##  forcats       * 0.5.2      2022-08-19 [1] CRAN (R 4.2.1)
##  foreign         0.8-82     2022-01-13 [4] CRAN (R 4.1.2)
##  Formula         1.2-4      2020-10-16 [1] CRAN (R 4.2.0)
##  fs              1.5.2      2021-12-08 [1] CRAN (R 4.2.0)
##  future          1.30.0     2022-12-16 [1] CRAN (R 4.2.1)
##  future.apply    1.10.0     2022-11-05 [1] CRAN (R 4.2.1)
##  gargle          1.2.1      2022-09-08 [1] CRAN (R 4.2.1)
##  generics        0.1.3      2022-07-05 [1] CRAN (R 4.2.1)
##  ggforce       * 0.4.1      2022-10-04 [1] CRAN (R 4.2.1)
##  ggplot2       * 3.4.0      2022-11-04 [1] CRAN (R 4.2.1)
##  globals         0.16.2     2022-11-21 [1] CRAN (R 4.2.1)
##  glue            1.6.2      2022-02-24 [1] CRAN (R 4.2.0)
##  googledrive     2.0.0      2021-07-08 [1] CRAN (R 4.2.0)
##  googlesheets4   1.0.1      2022-08-13 [1] CRAN (R 4.2.1)
##  gridExtra       2.3        2017-09-09 [1] CRAN (R 4.2.0)
##  gtable          0.3.1      2022-09-01 [1] CRAN (R 4.2.1)
##  haven           2.5.1      2022-08-22 [1] CRAN (R 4.2.1)
##  highr           0.10       2022-12-22 [1] CRAN (R 4.2.1)
##  Hmisc           4.7-2      2022-11-18 [1] CRAN (R 4.2.1)
##  hms             1.1.2      2022-08-19 [1] CRAN (R 4.2.1)
##  htmlTable       2.4.1      2022-07-07 [1] CRAN (R 4.2.1)
##  htmltools       0.5.4      2022-12-07 [1] CRAN (R 4.2.1)
##  htmlwidgets     1.6.1      2023-01-07 [1] CRAN (R 4.2.1)
##  httr            1.4.4      2022-08-17 [1] CRAN (R 4.2.1)
##  interp          1.1-3      2022-07-13 [1] CRAN (R 4.2.1)
##  jpeg            0.1-10     2022-11-29 [1] CRAN (R 4.2.1)
##  jsonlite        1.8.4      2022-12-06 [1] CRAN (R 4.2.1)
##  knitr         * 1.41       2022-11-18 [1] CRAN (R 4.2.1)
##  labeling        0.4.2      2020-10-20 [1] CRAN (R 4.2.0)
##  lattice         0.20-45    2021-09-22 [4] CRAN (R 4.2.0)
##  latticeExtra    0.6-30     2022-07-04 [1] CRAN (R 4.2.1)
##  lifecycle       1.0.3      2022-10-07 [1] CRAN (R 4.2.1)
##  listenv         0.9.0      2022-12-16 [1] CRAN (R 4.2.1)
##  lme4            1.1-29     2022-04-07 [1] CRAN (R 4.2.0)
##  lpSolve         5.6.18     2023-02-01 [1] CRAN (R 4.2.2)
##  lubridate       1.9.0      2022-11-06 [1] CRAN (R 4.2.1)
##  magrittr        2.0.3      2022-03-30 [1] CRAN (R 4.2.0)
##  MASS          * 7.3-58.2   2023-01-23 [4] CRAN (R 4.2.2)
##  Matrix        * 1.5-3      2022-11-11 [1] CRAN (R 4.2.1)
##  mediation     * 4.5.0      2019-10-08 [1] CRAN (R 4.2.2)
##  minqa           1.2.5      2022-10-19 [1] CRAN (R 4.2.1)
##  modelr          0.1.10     2022-11-11 [1] CRAN (R 4.2.1)
##  munsell         0.5.0      2018-06-12 [1] CRAN (R 4.2.0)
##  mvtnorm       * 1.1-3      2021-10-08 [1] CRAN (R 4.2.1)
##  nlme            3.1-162    2023-01-31 [4] CRAN (R 4.2.2)
##  nloptr          2.0.3      2022-05-26 [1] CRAN (R 4.2.0)
##  nnet            7.3-18     2022-09-28 [4] CRAN (R 4.2.1)
##  parallelly      1.34.0     2023-01-13 [1] CRAN (R 4.2.1)
##  patchwork     * 1.1.2      2022-08-19 [1] CRAN (R 4.2.1)
##  pillar          1.8.1      2022-08-19 [1] CRAN (R 4.2.1)
##  pkgconfig       2.0.3      2019-09-22 [1] CRAN (R 4.2.0)
##  png             0.1-8      2022-11-29 [1] CRAN (R 4.2.1)
##  polyclip        1.10-4     2022-10-20 [1] CRAN (R 4.2.1)
##  propr         * 4.2.6      2019-12-16 [1] CRAN (R 4.2.1)
##  purrr         * 1.0.1      2023-01-10 [1] CRAN (R 4.2.1)
##  R6              2.5.1      2021-08-19 [1] CRAN (R 4.2.0)
##  RColorBrewer    1.1-3      2022-04-03 [1] CRAN (R 4.2.0)
##  Rcpp            1.0.9      2022-07-08 [1] CRAN (R 4.2.1)
##  readr         * 2.1.3      2022-10-01 [1] CRAN (R 4.2.1)
##  readxl          1.4.1      2022-08-17 [1] CRAN (R 4.2.1)
##  reprex          2.0.2      2022-08-17 [1] CRAN (R 4.2.1)
##  rlang           1.0.6      2022-09-24 [1] CRAN (R 4.2.1)
##  rmarkdown       2.20       2023-01-19 [1] CRAN (R 4.2.1)
##  rpart           4.1.19     2022-10-21 [4] CRAN (R 4.2.1)
##  rstudioapi      0.14       2022-08-22 [1] CRAN (R 4.2.1)
##  rvest           1.0.3      2022-08-19 [1] CRAN (R 4.2.1)
##  sandwich      * 3.0-2      2022-06-15 [1] CRAN (R 4.2.2)
##  scales          1.2.1      2022-08-20 [1] CRAN (R 4.2.1)
##  sessioninfo     1.2.2      2021-12-06 [1] CRAN (R 4.2.0)
##  stringi         1.7.12     2023-01-11 [1] CRAN (R 4.2.1)
##  stringr       * 1.5.0      2022-12-02 [1] CRAN (R 4.2.1)
##  survival        3.4-0      2022-08-09 [4] CRAN (R 4.2.1)
##  tibble        * 3.1.8      2022-07-22 [1] CRAN (R 4.2.1)
##  tidyr         * 1.2.1      2022-09-08 [1] CRAN (R 4.2.1)
##  tidyselect      1.2.0      2022-10-10 [1] CRAN (R 4.2.1)
##  tidyverse     * 1.3.2      2022-07-18 [1] CRAN (R 4.2.1)
##  timechange      0.2.0      2023-01-11 [1] CRAN (R 4.2.1)
##  Tjazi         * 0.1.0.0    2023-04-26 [1] Github (thomazbastiaanssen/Tjazi@91f5c82)
##  tweenr          2.0.2      2022-09-06 [1] CRAN (R 4.2.1)
##  tzdb            0.3.0      2022-03-28 [1] CRAN (R 4.2.0)
##  utf8            1.2.2      2021-07-24 [1] CRAN (R 4.2.0)
##  vctrs           0.5.1      2022-11-16 [1] CRAN (R 4.2.1)
##  volatility    * 0.0.0.9000 2022-05-25 [1] Github (thomazbastiaanssen/Volatility@c1f50bf)
##  withr           2.5.0      2022-03-03 [1] CRAN (R 4.2.0)
##  xfun            0.36       2022-12-21 [1] CRAN (R 4.2.1)
##  xml2            1.3.3      2021-11-30 [1] CRAN (R 4.2.0)
##  yaml            2.3.6      2022-10-18 [1] CRAN (R 4.2.1)
##  zoo             1.8-12     2023-04-13 [1] CRAN (R 4.2.2)
## 
##  [1] /home/thomaz/R/x86_64-pc-linux-gnu-library/4.2
##  [2] /usr/local/lib/R/site-library
##  [3] /usr/lib/R/site-library
##  [4] /usr/lib/R/library
## 
## ------------------------------------------------------------------------------
\end{verbatim}